\documentclass[a4paper, 12pt]{article}
\pdfoutput=1
\usepackage[a4paper, inner=3cm, outer=3cm, top=3cm, bottom=3cm]{geometry}
\usepackage{cite}
\usepackage{epsfig,amssymb,euscript,xspace,xcolor,setspace,amsmath,mathtools,empheq,amsthm,graphicx,mathrsfs,float}
\usepackage{hyperref}
\usepackage{bbding}
\usepackage{xcolor}
\hypersetup{
    colorlinks=true,
    linkcolor=black,
    citecolor=blue
    } 
\usepackage{caption,subcaption} 
\usepackage[belowskip=-20pt,aboveskip=0pt]{caption}
\usepackage[skip=10pt]{caption} % example skip set to 2pt
\usepackage{comment}
\usepackage{wrapfig}

\usepackage[numbers,sort&compress]{natbib}
\usepackage{cite}
\newcommand\blfootnote[1]{%
  \begingroup
  \renewcommand\thefootnote{}\footnote{#1}%
  \addtocounter{footnote}{-1}%
  \endgroup
}

\makeatother

\def\Prho{{\mathcal P}_\rho}

\begin{document}
\baselineskip 24pt

\begin{center}

{\Large \bf  Celestial insights into the S-matrix bootstrap} 

\end{center}

\vskip .5cm
\medskip

\vspace*{4.0ex}

\baselineskip=18pt

\centerline{\large \rm  Sudip Ghosh$^{\omega}$\blfootnote{$^{\omega}$sudip112phys@gmail.com}, Prashanth Raman$^{z}$ \blfootnote{$^{z}$prashanth.raman108@gmail.com} and Aninda Sinha$^{\bar{z}}$\blfootnote{$^{\bar{z}}$asinha@iisc.ac.in}  }
\vspace*{4.0ex}

\centerline{\it  Centre for High Energy Physics, Indian Institute of Science,}
\centerline{\it~ C.V. Raman Avenue, Bangalore 560012, India.}

\vspace*{1.0ex}
\centerline{\it ~} 
\vspace*{5.0ex}
\centerline{\bf Abstract} \bigskip
We consider 2-2 scattering in four spacetime dimensions in Celestial variables. Using the crossing symmetric dispersion relation (CSDR), we recast the Celestial amplitudes in terms of crossing symmetric partial waves. These partial waves have spurious singularities in the complex Celestial variable, which need to be removed in local theories. The locality constraints (null constraints) admit closed form expressions, which lead to novel bounds on partial wave moments. These bounds allow us to quantify the degree of low spin dominance(LSD) for scalar theories. We study a new kind of positivity that seems to be present in a wide class of theories. We prove that this positivity arises only in theories with a spin-0 dominance. The crossing symmetric partial waves with spurious singularities removed, dubbed as Feynman blocks, have remarkable properties in the Celestial variable, namely typically realness, in the sense of Geometric Function Theory (GFT). Using GFT techniques we derive non-projective bounds on Wilson coefficients in terms of partial wave moments.  
\vfill \eject
\baselineskip=18pt

\tableofcontents
\onehalfspacing

%\tableofcontents

\newpage

\section{Introduction} 
\label{sec:intro}
Over the last few years, it has been realized that combining the power of dispersion relations and crossing symmetry leads to powerful constraints on the low energy expansion of 2-2 scattering \cite{snowsmat1,snowsmat2}. At the same time, to understand scattering of massless particles in four spacetime dimensions, the program bearing the name of ``Celestial amplitudes'' \cite{Pasterski:2021raf, McLoughlin:2022ljp} has been an active area of research. 

Celestial amplitudes represent S-matrix elements in a basis where the external particles are in boost eigenstates. In this basis, $4$-$d$ scattering amplitudes manifestly transform as $2$-$d$ conformal correlation functions \cite{Pasterski:2016qvg,Pasterski:2017kqt,Banerjee:2018gce}. Due to this feature, celestial amplitudes have emerged as a central object of interest in the context of flat space holography, where recent developments on the connection between soft theorems and asymptotic symmetries suggest that the holographic dual of quantum gravity in asymptotically flat spacetimes is a $2$-$d$ celestial conformal field theory (CCFT) defined on the celestial sphere at null infinity \cite{Strominger:2017zoo, Kapec:2014opa, Kapec:2016jld}. 

The celestial formalism has led to several fascinating recent insights, particularly for scattering of massless particles (see \cite{Pasterski:2021raf, McLoughlin:2022ljp,Pasterski:2021rjz,Raclariu:2021zjz} for recent reviews). For example, soft theorems in gravity and gauge theories in $4$-dimensions have been shown to imply the existence of infinite dimensional current algebra symmetries acting on the $2$-$d$ celestial sphere \cite{Banerjee:2020zlg, Banerjee:2020vnt, Guevara:2021abz, Strominger:2021mtt}. These symmetries impose powerful constraints on the operator product expansion (OPE) in CCFT, which in turn is related to collinear limits of scattering amplitudes \cite{Fan:2019emx, Pate:2019lpp, Banerjee:2020kaa}. Quite remarkably, these infinite-dimensional celestial symmetries can be used to completely determine tree-level MHV amplitudes in Yang-Mills theory and Einstein gravity \cite{Banerjee:2020zlg, Banerjee:2020vnt}. In this paper, we wish to understand what insights one can obtain about the S-matrix bootstrap program using ideas from celestial amplitudes. In a companion paper \cite{toappear}, we will present what can be learnt about CCFT from the corresponding bulk effective field theory (EFT) using insights gained from the S-matrix bootstrap.  
 
A main recent development is the derivation of two-sided bounds on ratios of Wilson coefficients\footnote{Taylor expansion coefficients arising in the low energy expansion of 2-2 scattering amplitudes, which in turn are related to contact vertices in the effective action.} \cite{snowsmat1,snowsmat2}. The primary tool in this area of research has been the fixed-$t$ dispersion relation. Since 2-2 scattering is a function of the two Mandelstam invariants $s,t$, historically much attention has focused on a dispersion relation where one of these variables (typically $t$) is held fixed. In the case of scattering of identical particles, a penalty that one has to pay is the loss of crossing symmetry which has to be imposed as a constraint. Such constraints have been dubbed as ``null constraints" in \cite{Tolley:2020gtv,sch1}. Using these constraints, and linear programming, one numerically finds two-sided bounds on Wilson coefficients. In \cite{efthedron} a geometric picture was put forward where it was argued that as a consequence of the constraints arising from locality and unitarity the space of Wilson coefficients was forced to lie inside a geometric region called the  EFThedron. 

In the early 1970s, Auberson and Khuri had looked at a dispersion relation with manifest crossing symmetry (CSDR). This line of research lay dormant for many years. Recently this dispersion relation was resurrected in \cite{Sinha:2020win,Gopakumar:2021dvg}. Since there is inbuilt crossing symmetry at the onset, the penalty one pays to have a dispersion relation is the loss of manifest locality. Namely one finds spurious poles in the partial waves. Cancellation of these spurious poles is needed to have a local low energy expansion. This can only happen after summing over spins. The role of ``null constraints" is played by these ``locality constraints" in this program. In \cite{Sinha:2020win, Gopakumar:2021dvg}, the equivalence of the two sets of constraints was shown. The partial waves in the CSDR with spurious singularities were termed as ``Dyson blocks'' in \cite{Sinha:2020win}. There is another version of partial waves, which look closer in spirit to Feynman diagrams, which are singularity free and resemble exchange Feynman diagrams with specific contact diagrams. These were called ``Feynman blocks'' in \cite{Sinha:2020win}. As we will see, these Feynman blocks in the Celestial variables have remarkable properties.

One of the main advantages of working with the CSDR is that it leads to a fascinating connection with an area of mathematics called Geometric Function Theory (GFT) \cite{hsz, rs1, rs2,Zahed:2021fkp}. The origin of two sided bounds on Wilson coefficients gets related to the famous Bieberbach conjecture (de Branges' theorem). The main property of the amplitude that enables this connection is what is called ``typically real''-ness or ``Herglotz''. A function $f(z)$ is typically real if it satisfies $Im f(z) Im z> 0$ whenever $Im z \neq 0$. If the function is regular inside the unit disk then the Taylor expansion coefficients of the function satisfy two-sided bounds called Bieberbach-Rogosinski bounds \cite{rs1}. The function is also allowed to have simple poles on the real axis. When this happens, the two-sided bounds get modified to the so-called Goodman bounds where the gap between the origin and the nearest pole controls the two-sided bounds. These mathematical facts are reviewed in \cite{rs1}. 

A feature of the crossing symmetric dispersive representation of the amplitude is that it involves a kernel (to be reviewed in section \ref{subsec:csdrreview}) which bears resemblance with tree-level $\phi^3$ theory. $4$-point celestial amplitudes and their conformal block decomposition for tree-level $\phi^3$ theory involving massive scalar exchange have recently been studied in \cite{Lam:2017ofc,Nandan:2019jas,Atanasov:2021cje}.  It is therefore naturally tempting to build on these works using the CSDR.  The above mentioned kernel in CSDR is also dressed with the Legendre (Gegenbauer) polynomials, which carry information about spins in the partial wave expansion of the amplitude. For spin-0, the results of \cite{Lam:2017ofc,Nandan:2019jas,Atanasov:2021cje}  for the celestial amplitude can be readily imported.  With some more effort, we will be able to calculate the $4$-point celestial amplitude for any spin involved in the partial wave decomposition of the momentum space amplitude using the CSDR.

In this paper, one of our main objective is to explain what insights can be obtained for EFTs using CCFT techniques. For this purpose, we will consider $2$-$2$ scattering of massless particles and write the Mandelstam variables in terms of the celestial variable, $z$, as follows 
\begin{equation}
    s=\omega^2,\quad t=-\omega^2 z,\quad u=-\omega^2(1-z)\,.
\end{equation}

For fixed $z$, this choice of variables enables one to study fixed-angle scattering. As reviewed in section \ref{subsec:mellincsdr}, the celestial amplitude is obtained as a Mellin transform of the four dimensional scattering amplitude. The Mellin variable is $\beta$. 
We will show how to repackage the information about the null/locality constraints systematically in the celestial basis. Next we will examine the properties of the Feynman blocks in the celestial variable. Specifically, we will be interested in the residues of the celestial amplitude at $\beta=-2n$, i.e., negative even integers, since these contain information about the low-energy expansion coefficients in the momentum space amplitude \cite{Arkani-Hamed:2020gyp, Chang:2021wvv}. For each $n$, we can write an explicit expression for the amplitude in terms of a sum over Feynman blocks. Quite remarkably, we will find that beyond a certain critical spin $J=J_T$,  all the Feynman blocks are typically real polynomials! This enables us to put two-sided bounds on the truncated partial wave sum $J<J_T$ using polynomial analogues of the Bieberbach-Rogosinski bounds for typically real functions which we call Suffridge bounds \eqref{suffridgedistortion}. Furthermore, this leads to novel two sided bounds on the Wilson coefficients themselves in terms of the $J<J_T$ partial wave moments. 

Let us now give a brief overview of the key results. %{\bf SG, PR please fill this part in an itemized format.} 
We have been able to: 
\begin{itemize}
    \item  Show that there is a new kind of positivity exhibited by amplitudes in terms of a variable $\rho$ which is related to celestial variable $z$ as $\rho=-1-2z(z-1)$.
    
    \item Obtain a representation for the $4$-point celestial amplitude of massless scalars using the  crossing symmetric dispersive representation of momentum space amplitude  \eqref{mtildebetaz1} for generic $\beta$ and by specializing to $\beta=-2n, ~n \in  \mathbb{Z}_+$ relevant for low-energy physics \eqref{betaresrhovar}, systematically analyze the implications of locality constraints \eqref{locconstr}.
    
    \item Obtain bounds on partial wave moments as a direct consequence of the above mentioned locality constraints \eqref{locconstr}, using which we quantify the phenomenon of low spin dominance (LSD) and argue that the $\rho$-positivity is tied to spin-0 dominance \eqref{lsd}.
    
    \item Show that as a function of $\rho$ the Feynman blocks for large enough spins are typically-real polynomials. Using this we have been able to put non-projective bounds on the low energy Wilson coefficients \eqref{n4bounds},\eqref{pw51} in terms a few low spin partial wave moments. 
    
    \item Obtain bounds for the case with graviton exchange in the amplitude by using Goodman bounds for typically real functions in the variable $\tilde{\rho}=\rho+1$.
\end{itemize}

A question worth asking at this point is if one could have obtained these results without appealing to CCFTs. The key player in our story is the Celestial variable $\rho$ and GFT methods relying on typically-realness in this variable. It is unclear why one would be interested in analysing such properties in this variable without having the motivation to understand CCFTs, which is why we feel that the CCFT formalism has been the key player leading to the S-matrix insights obtained in this paper.   

The paper is organized as follows. %{\bf  please fill}
In section 2, we begin by introducing the celestial inspired $\rho$-variable, in which known amplitudes curiously seem to exhibit a hitherto unknown kind of positivity. In section 3, by starting with the CSDR we obtain a representation of the celestial amplitude for generic $\beta$. In section 4, by specializing to $\beta=-2n$, we analyze the locality constraints which imply certain bounds on partial wave moments, LSD and a connection between $\rho$-positivity and spin-0 dominance. In section 5, we show that there is a connection between the Feynman blocks and typically-real polynomials in the unit disk $|\rho|<1$ and use techniques from GFT to obtain two sided bounds on low energy Wilson coefficients $\mathcal{W}_{pq}$ in terms of lower spin partial waves. We conclude in section 6 with a discussion on the possible future directions of interest. The appendices supplement the material in the main text with proofs, closed form expressions and tables of data.

%It is naturally tempting to build on \cite{Atanasov:2021cje} using the CSDR. First, the crossing kernel (to be reviewed below) has the resemblance with the tree-level $\phi^3$ theory. In the CSDR, the kernel is dressed with the Legendre (Gegenbauer) polynomials, which carry information about the spin. For spin-0, the results of \cite{Atanasov:2021cje}  can be readily imported. With some more effort, we will be able to calculate for any spin.
%%%%%%%%%%%%%%%%%%%%%%%%%%%%%%%%%%%%%%%%%%%%%%%%%%%%%%%%%%%%%%%%%%%%%%%%%%%%%%%
\section{Celestial insight 1: A curious observation}
\label{sec:newpositivity}

%{\bf AS: I have fixed minus sign convention as we discussed, please check. }

In this section, we wish to point out an interesting feature of the low energy expansion of 2-2 scattering in many theories. We will start with string theory. Consider the following two fully crossing symmetric amplitudes \cite{nimayutinstring}.

\begin{equation}
    \begin{split}\label{CBamplitude}
        \mathcal{M}_{CB}(s,t)=  \frac{\Gamma(-s-1)\Gamma(-t-1)\Gamma(-u-1)}{\Gamma(2+s)\Gamma(2+t)\Gamma(2+u)}\,, \quad s+t+u=-4
    \end{split}
\end{equation}

\begin{equation} \label{type2}
    \begin{split}
        M_{II}(s,t) = -x^2 \mathcal{M}_{II}(s,t)\equiv x^2\frac{\Gamma(-s)\Gamma(-t)\Gamma(-u)}{\Gamma(1+s)\Gamma(1+t)\Gamma(1+u)}\,,\quad s+t+u=0
    \end{split}
\end{equation}

Here we have defined $x= -(st+tu+su)$. The first amplitude is the 2-2 tree level scattering of tachyons in closed bosonic string theory while the second one is the 2-2 tree level scattering of dilatons in type-II string theory. For type-II, we can also consider the 2-2 graviton scattering amplitude ${\mathcal R}^4 \mathcal{M}_{II}(s,t)$. 

\vskip 3pt
We wish to expand both amplitudes in a manifestly crossing symmetric manner. To this effect we will introduce
\begin{eqnarray}
s_1&=&s+\frac{4}{3}\,,\quad s_2=t+\frac{4}{3}\,, \quad s_3=u+\frac{4}{3}\,, \quad{\rm for~ }\mathcal{M}_{CB}\\
s_1&=& s\,,\quad s_2=t\,,\quad s_3=u\,, \quad {\rm for~} \mathcal{M}_{II}\,.
\end{eqnarray}

In both cases $s_1+s_2+s_3=0$. Now we introduce the celestial variables
\begin{equation}
s_1=\omega^2\,,\quad s_2=-\omega^2 z\,,\quad s_3=-\omega^2(1-z)\,.    
\end{equation}

Further for later convenience, we introduce
\begin{equation}
    \rho=-1-2z^2+2z\,.
\end{equation}

The relation between the $\rho$ variable and the $z$ variable is indicated in the figure.
\begin{figure}[ht]
\centering
  \includegraphics[width=0.5\linewidth]{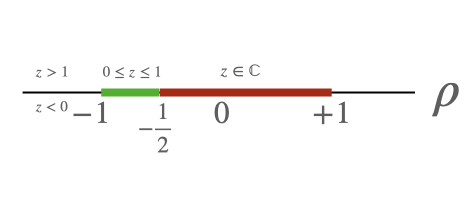}
  \caption{The $\rho$-variable. We have marked the range of interest. Green indicates physical region while red indicates unphysical region.}
  \label{fig:pic1a}
\end{figure}

\vskip 4pt
In passing, we also note the following useful relations ($s_1+s_2+s_3=0$):
\begin{eqnarray}
x&\equiv &-(s_1 s_2+s_1 s_3+s_2 s_3)=\frac{\omega^4}{2}(1-\rho)\,,\\
y&\equiv &-s_1 s_2 s_3=-\frac{\omega^6}{2}(1+\rho)\,.
\end{eqnarray}

Then expanding the amplitudes around $\omega^2=0$ (i.e., the crossing symmetric point), we find\footnote{One cross-check about the overall sign is that if we write the expansion as $\sum {\mathcal W}_{pq}x^p y^q$, then using unitarity, one can show that $\mathcal{W}_{n,0}\geq 0$. In other words, the coefficients of $\omega^{4n}(1-\rho)^n$ are guaranteed to be positive.}
\begin{equation}
\begin{split}\label{cbtype2}
 & \mathcal{M}_{CB}   \approx 7.74+27.22  \omega^4(1-\rho)+121 \omega^6(1+\rho)+121.56 \omega^8 (1-\rho)^2+911.18\omega^{10}(1-\rho^2) \\
& \hspace{1.4cm} +546.77 \omega^{12}\left[(1-\rho)^3+3(1+\rho)^2\right]+O(\omega^{14})\,,\\ \\ 
&\mathcal{M}_{II}-\frac{2}{\omega^6(1+\rho)}\approx
2.40+1.04\omega^4(1-\rho)+1.44\omega^6(1+\rho)+0.50\omega^8(1-\rho)^2 \\
& \hspace{3.4cm} +1.25 \omega^{10}(1-\rho^2)
+0.25\omega^{12}\left[(1-\rho)^3+2.98(1+\rho)^2\right]+O(\omega^{14})\,.
\end{split}
\end{equation}

For type II, we have put the graviton pole on the left. 

Now consider two more cases. First, the run-of-the-mill $\phi^2\psi$ theory at tree level where we are scattering massless $\phi$ which exchanges a massive $\psi$ at tree level. The amplitude for this is:

\begin{eqnarray}
\mathcal{M}_{\phi^2 \psi}&=& g^2\left(\frac{1}{m^2-s}+\frac{1}{m^2-t} +\frac{1}{m^2-u}  \right) \nonumber\\
&=&\frac{3 g^2}{m^2}+\frac{g^2 \omega^4}{m^6}(1-\rho)+\frac{3 g^2 \omega^6}{2 m^8}(1+\rho)+\frac{g^2\omega^8}{2m^{10}}(1-\rho)^2+\frac{5 g^2\omega^{10}}{4 m^{12}}(1-\rho^2)\nonumber\\&+&\frac{g^2 \omega^{12}}{4m^{14}}\left[(1-\rho)^3+3(1+\rho)^2\right]+O(\omega^{14})\,.
\end{eqnarray}

Finally consider the theory at one-loop which is given in terms of the Appell $F_3$ \cite{davdychev}
\begin{eqnarray}
\mathcal{M}_{\phi^2 \psi}&=& \frac{\pi^2}{6 m^4}\left( F_3 \left({{1,1,1,1}\atop{\frac{5}{2}}}; \frac{s}{4 m^2},\frac{t}{4 m^2} \right)+(s\rightarrow t, t\rightarrow u)+(s\rightarrow u, t\rightarrow s) \right) \nonumber \\
&=&\frac{\pi^2}{6 m^4}\left( \sum_{p,q=0}^{\infty} \frac{p!~q!}{\left(\frac{5}{2}\right)_{p+q}(4~m^2)^{p+q}}\left(s^p t^q+t^p u^q+u^p s^q\right)\right) \nonumber\\
&\approx& \frac{\pi^2}{6 m^4}\bigg(3+\frac{\omega^4}{40 m^4}(1-\rho)+\frac{\omega^6}{168 m^6}(1+\rho)+\frac{\omega^8}{2520 m^8}(1-\rho)^2+\frac{\omega^{10}}{5280 m^{10}}(1-\rho^2)\nonumber\\
&+& \frac{\omega^{12}}{128128 m^{12}}\left[(1-\rho)^3+2.91(1+\rho)^2\right]\bigg)+O(\omega^{14})
\end{eqnarray}

Now all of these expansions have the following startling feature in common. 

\smallskip
{\it All these expansions up to any fixed order in $\omega$ are positive polynomials in $\rho$ in the interval $\rho\in(-1,1)$. In order to be concise, we will refer to this positivity as $\Prho$.}
\smallskip

A positive polynomial $p(x)$ on an interval $(a,b)$ is one that is $p(x) \ge 0 ~\forall~ x\in(a,b)$. A nice characterization of such polynomials \cite{powers} on $(-1,1)$ is that they can be expanded in terms of the so called {\it Bernstein basis} $p(x)=\sum_{i=0}^m c_{i}(1+x)^{m-i} (1-x)^i$ such that $c_{i,j}\ge 0$ however we may need $m\ge d$ where d is the degree of the polynomial. The smallest $m$ such that for $n\ge m$ guarantees $c_i \ge 0$ is called the Bernstein degree of the polynomial \cite{powers}. The Bernstein degree requires knowledge of the maximum and minimum values of the polynomial $p(x)$. In appendix \ref{sec:positivityproofs}, we will derive these positivity properties directly using the known expressions for the amplitudes. 

 Now some of the positivity features can be explained quite straightforwardly using a dispersion relation. For instance, the coefficient of the $\omega^4(1-\rho)$ term can be shown to be positive using partial wave unitarity. The full positivity in the $\rho\in(-1,1)$ interval however is harder to explain. One of the main purposes of this paper is to find analytic conditions under which such positivity can hold. Our main tool will be to use the crossing symmetric dispersion relation (CSDR) \cite{Auberson:1972prg, Sinha:2020win} which we will review next.
%%%%%%%%%%%%%%%%%%%%%%%%%%%%%%%%%%%%%%%%%%%%%%%%%%%%%%%%%%%%%%%%%%%%%%%%%%%%%%
\section{Essential technicalities: Dispersion relations}
As mentioned in the introduction, our focus in this paper will be the use of the crossing symmetric dispersion relation (CSDR). Many of the analytic properties will be transparent using the CSDR \footnote{It should be possible to use the fixed-$t$ dispersion relations to find numerical evidence for these properties but we will leave this as an open problem.}. We begin with a lightning review of the CSDR. For further details, we refer the reader to
\cite{Auberson:1972prg,Sinha:2020win}.

\subsection{CSDR: A quick review}
\label{subsec:csdrreview}

Consider $\mathbf{M}(s,t)$ to be the $2$-$2$ scattering amplitude of identical massless scalars in four spacetime dimensions. $\mathbf{M}(s,t)$ admits a crossing symmetric dispersive representation given by \cite{Auberson:1972prg,Sinha:2020win}
\begin{equation}
\label{csdr}
\begin{split}
\mathbf{M}(s,t) = c_{0}+  \frac{1}{\pi} \int_{\delta_{0}}^{\infty} \frac{ds'}{s'} \hspace{0.1cm} \mathcal{A}(s', a) H(s'; s,t,u)
\end{split}
\end{equation}

where 
\begin{equation}
\label{adef}
\begin{split}
a = \frac{s t u}{st + t u + u s}\equiv \frac{y}{x}\,.
\end{split}
\end{equation}
Here $\delta_0$ is the location of the cut (or in the case of string theory, the first massive pole) and $c_0=\mathbf{M}(0,0)$, which arises as we have assumed two subtractions while writing down the dispersion relation \cite{Auberson:1972prg,Sinha:2020win}. 
$\mathcal{A}(s, a)$  is the $s$-channel discontinuity of the amplitude and $H(s'; s,t,u) $ denotes the following crossing symmetric kernel
\begin{equation}
\label{kernel}
\begin{split}
H(s'; s,t,u) = \frac{s}{s'-s} + \frac{t}{s'-t}+ \frac{u}{s'-u}\,.
\end{split}
\end{equation}

The discontinuity $\mathcal{A}(s, a)$ can be expanded in terms of Legendre polynomials as
\begin{equation}
\label{absp}
\begin{split}
\mathcal{A}(s', a) = 32\pi  \sum_{J=0}^{\infty} (2J+1) \ \alpha_{J}(s') \ P_{J}\left( \sqrt{\frac{s'+3a}{s'-a}}\right)\,.
\end{split}
\end{equation}
where $\alpha_{J}(s')$ are the partial wave coefficients. In the sum over spins in \eqref{absp}, only even spins contribute since we are considering here the amplitude for identical scalars. The conventions are chosen so that unitarity leads to $0\leq \alpha_J(s)\leq 1$. 

Now the nontrivial form of the argument of the Legendre polynomial is to be noted. When the theory is gapped, it is known \cite{Auberson:1972prg}, that the partial wave expansion converges over a range of the parameter $a$, which allows for Taylor expanding around $a\sim 0$. Since $a$ involves inverse powers of $x$, this would lead to negative powers of $x$ in a particular partial wave. In a local theory, these inverse powers of $x$ should be absent. This means that when we sum over the spins, such inverse powers should cancel. This leads to what we call ``locality'' constraints. In \cite{Sinha:2020win}, it was shown that these are equivalent to the so-called ``null constraints'' which arise on imposing crossing symmetry on the fixed-$t$ dispersion relation \cite{sch1, tolley1}. 

%In the context of EFTs, the lower limit  $\mu^{2}$  of the integral in \eqref{csdr} denotes the EFT scale. 

%%%%%%%%%%%%%%%%%%%%%%%%%%%%%%%%%%%%%%%%%%%%%%%%%%%%%%%%%%%%%%%%%%%%%%%%%%%%
\subsection{Dispersion relation in celestial basis}
\label{subsec:mellincsdr}

%{\bf SG, PR: This section needs polishing. SG wanted to remove much of the stuff to an appendix. Also lengthy formulas which makes our eyes typically gloss over should be moved to an appendix as well.}

In this section we consider the $4$-point celestial amplitude for identical massless scalars in four spacetime dimensions and evaluate it using the crossing symmetric dispersive representation of the momentum space amplitude given in section \ref{subsec:csdrreview}.  In order to write down the celestial amplitude, the null four-momenta of the external particles can be parametrized as 
\begin{equation}
\label{nullmompar}
\begin{split}
p^{\mu}_{k} = \epsilon_{k} \omega_{k} (1+ z_{k} \bar{z}_{k}, z_{k}+ \bar{z}_{k}, -i (z_{k}-\bar{z}_{k}), 1-z_{k}\bar{z}_{k}), \quad k=1,2,3,4
\end{split}
\end{equation}

where $\epsilon_{k} =\pm 1$ for an outgoing (incoming) particle. $\omega_{k}$ is the energy of the $k$-th particle. $(z_{k},\bar{z}_{k})$ specify the directions of null-momenta of the asymptotic states in the S-matrix and hence can be regarded as stereographic coordinates on the $2$-$d$ celestial sphere. Throughout the rest of this paper, we take $\epsilon_{1}=\epsilon_{2}= -1$ and $\epsilon_{3}=\epsilon_{4}=1$ corresponding to particles $(1,2)$ incoming and $(3,4)$ outgoing. 

\vskip 4pt
The $4$-point celestial amplitude  is then given by
\begin{equation}
\label{csamp}
\begin{split}
\mathcal{M}(\Delta_{i},  z_{i},\bar{z}_{i}) & =  \int_{0}^{\infty} \prod_{i=1}^{4} d\omega_{i} \ \omega_{i}^{\Delta_{i}-1} \ \mathbf{M}(\omega_{i},z_{i},\bar{z}_{i}) \ \delta^{(4)}\left(\sum_{i=1}^{4}p^{\mu}_{i}(\omega_{i},z_{i},\bar{z}_{i})\right)
\end{split}
\end{equation}

where $\mathbf{M}(\omega_{i},z_{i},\bar{z}_{i})$ is the momentum space amplitude with the external momenta parametrized as in \eqref{nullmompar}. Under the action of the Lorentz group which acts as $SL(2,C)$ on the $(z_{i},\bar{z}_{i})$ variables, the celestial amplitude $\mathcal{M}(\Delta_{i},  z_{i},\bar{z}_{i})$ transforms as a $4$-point correlation function of quasi-primary operators with scaling dimension $\Delta_{i}$ in a $2$-$d$ CFT which in this context is referred to as Celestial CFT (CCFT).   

Now $\mathcal{M}(\Delta_{i},  z_{i},\bar{z}_{i})$ can be further expressed as \cite{Arkani-Hamed:2020gyp}\footnote{See Appendix \ref{sec:csampreview} for a review of the derivation of \eqref{csamp1}.}
\begin{equation}
\label{csamp1}
\begin{split}
& \mathcal{M}\left( \Delta_{i}, z_{i},\bar{z}_{i} \right)  =  \prod_{i<j}^{4} ( z_{ij}\bar{z}_{ij})^{\frac{1}{2}\left(\frac{\Delta}{3}-\Delta_{i}-\Delta_{j}\right)} \ 2^{-\beta-2}  \ |z(1-z)|^{\frac{(\beta+4)}{6}} \ \delta(z-\bar{z}) \  \widetilde{\mathbf{M}}(\beta, z) 
\end{split}
\end{equation}

where 
\begin{equation}
\label{KandXdef}
\begin{split}
& \Delta= \sum_{i=1}^{4}\Delta_{i} ; \quad \beta= \Delta-4
\end{split}
\end{equation}

$z,\bar{z}$ denote the cross ratios
\begin{equation}
\label{crs}
\begin{split}
z =\frac{z_{13}z_{24}}{z_{12}z_{34}}; \quad \bar{z} = \frac{\bar{z}_{13}\bar{z}_{24}}{\bar{z}_{12}\bar{z}_{34}}
\end{split}
\end{equation}

and 
\begin{equation}
\label{Mtildedef}
\begin{split}
\widetilde{\mathbf{M}}(\beta, z)   = \int_{0}^{\infty}  d\omega \ \omega^{\beta-1} \ \mathbf{M}(\omega^{2}, -  z \omega^{2}) 
\end{split}
\end{equation}

$\widetilde{\mathbf{M}}(\beta, z)$ is the Mellin transform of the of $2$-$2$ momentum space amplitude $\mathbf{M}(s,t)$ where the Mandelstam invariants have been parametrized as
\begin{equation}
\label{stuzvar}
\begin{split}
s = \omega^{2}; \quad t= -z\omega^{2}; \quad u= (z-1) \omega^{2}
\end{split}
\end{equation}

Here $z = - t/s $ is related to the scattering angle $\theta$ in the $s$-channel via $z= \frac{1}{2}(1-\cos \theta)$. For physical $s$-channel kinematics we thus have $z \in [0,1]$. In this paper one of the central objects of interest is $\widetilde{\mathbf{M}}(\beta, z)$. Since the kinematic prefactors in \eqref{csamp1} will be irrelevant for our purposes here, we will refer to $\widetilde{\mathbf{M}}(\beta, z)$ simply as the celestial or Mellin amplitude in the rest of this paper.

Let us now determine the Mellin amplitude using the representation of $\mathbf{M}(s,t)$ given by the crossing symmetric dispersion relation \eqref{csdr}. For this, we use the partial wave expansion \eqref{absp} and also apply the celestial parametrization \eqref{stuzvar}. The Mellin integral over $\omega$ can then be performed and we obtain 
\begin{equation}
\label{mtildebetaz1}
\begin{split}
& \widetilde{\mathbf{M}}(\beta, z)   \\
&=  2\pi c_{0} \delta(\beta)   +  \frac{  \pi }{2 \sin\left( \frac{\pi \beta}{2}\right)}   \sum_{J=0}^{\infty} (2J+1) \  \widetilde{\alpha}_{J}(\beta,\delta_{0})    \bigg[  e^{- i\pi \beta/2}   P_{J}\left( 1-2 z \right) + z^{-\beta/2} P_{J}\left(  \frac{z-2}{z}  \right)  \\
& +   (1-z)^{-\beta/2} P_{J}\left(  \frac{z+1}{z-1}  \right) +   (z(1-z))^{-\beta/2-J} (z^{2}-z+1)^{\beta/2+3}  \mathcal{Q}_{J}(\beta,z) \bigg] 
\end{split}
\end{equation}
where $\widetilde{\alpha}_{J}(\beta,\delta_{0})$ is given by %\textcolor{green}{\bf PR: They are moments only when $\beta=-2n$ right ?}
\begin{equation}
\label{alphatdef}
\begin{split}
\widetilde{\alpha}_{J}(\beta,\delta_{0}) = 32\int_{\delta_{0}}^{\infty} ds' \ s'^{\beta/2-1}   \ \alpha_{J}(s') \,,
\end{split}
\end{equation}
and are partial-wave moments for $\beta=-2n.$
$\mathcal{Q}_{J}(\beta,z)$ is a polynomial in $\beta, z$ and is given by
\begin{equation}
\label{Qjdef}
\begin{split}
  \mathcal{Q}_{J}(\beta, z) = (z(1-z))^J (z^2-z+1)^{-3} \sum\limits_{i=1}^3 \mathcal{R}_{J}(\beta,x_{i})
\end{split}
\end{equation}

where
\begin{equation}
\label{Rjdef}
\begin{split}
  \mathcal{R}_{J}(\beta, x_{i}) = \frac{e^{- i \pi \beta/2}}{\left(\frac{J}{2}-1\right)!} \  \frac{d^{J/2-1}}{dx^{J/2-1}}\bigg[ \frac{x^{\beta/2} (1+x)^{J/2}}{(x-x_{i})} P_{J}\left( \sqrt{\frac{1-3x}{1+x}}\right) \bigg] \bigg|_{x=-1}
\end{split}
\end{equation}

with $x_{1} = -z(z-1)(z^{2}-z+1)^{-1}, \ x_{2} = (z-1)(z^{2}-z+1)^{-1}, \ x_{3} = -z(z^{2}-z+1)^{-1}$. 

\vskip 4pt

\begin{figure}[h]
\centering
  \includegraphics[width=0.5\textwidth]{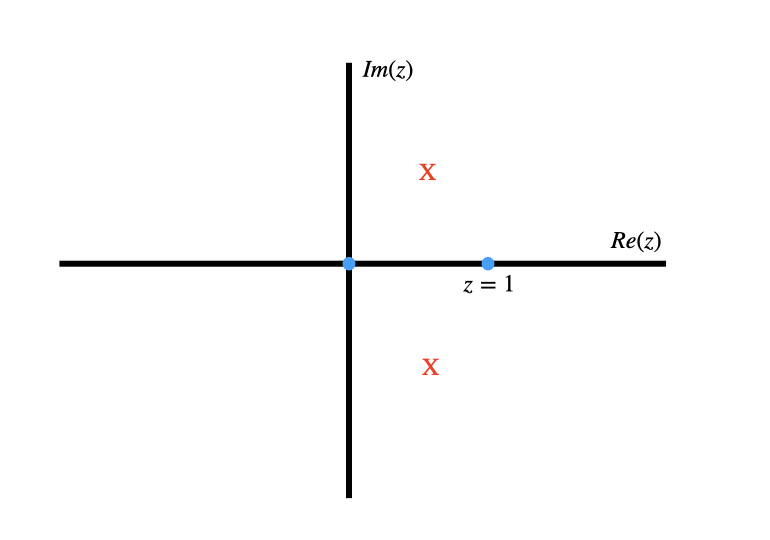}
  \caption{The analytic structure for each spin-$J$ partial wave. The blue circles are potential singularities at $z=0,1$ while the red crosses are potential singularities at $z^2-z+1=0$.}
  \label{fig:pic2a}
\end{figure} 

\vskip 10pt
We refer the reader to the Appendix, section \ref{sec:formulas} where a closed form expression for $\mathcal{Q}_{J}(\beta,z)$ is given. For illustrative purposes, we note below some explicit examples of $\mathcal{Q}_{J}(\beta,z)$ for spin $J=0,2,4,6$. 

\begin{equation}
\label{Qjdef}
\begin{split}
  & \mathcal{Q}_{2}(\beta, z) = 0, \quad \mathcal{Q}_{2}(\beta, z)=-6, \quad \mathcal{Q}_{4}(\beta, z)  = \frac{35}{4} (1+\rho )^3+ \frac{1}{4} (35 \beta+190) (1+\rho)^2-70, \\
  & \mathcal{Q}_{6}(\beta, z) = -\frac{21}{16} (11 \beta +57) (\rho +1)^5 -\frac{21}{32} \left(11 \beta ^2+136 \beta +424\right) (\rho +1)^4 \\
  & -\frac{21}{16} (132 \beta +508) (\rho +1)^3-\frac{21}{8} (44 \beta +564) (\rho +1)^2-\frac{231}{16}  (\rho +1)^6+2772 (\rho +1)+924
\end{split}
\end{equation}
where $\rho =-1-2z(z-1)$. The analytic structure in the complex-$z$ plane is indicated in the above figure  for $\beta \in -2 \mathbb{Z}$. The $z=0,1$ poles \footnote{For generic $\beta$ singularity structure is more complicated since from the expression \eqref{mtildebetaz1} it is manifest that there are branch cuts.} in each channel are cancelled for each $J$ when the crossing symmetric combination is used. The $z^2-z+1=0$ or $\rho=1$ singularities are what will lead to locality constraints discussed below.

The expression for $\widetilde{\mathbf{M}}(\beta, z)$ given by \eqref{mtildebetaz1} provides a representation of the $4$-point celestial amplitude for massless scalars in terms of the partial wave expansion of the momentum space amplitude. In the following sections we will primarily focus on the residues at the poles of $\widetilde{\mathbf{M}}(\beta, z)$ with respect to the parameter $\beta$ for negative integer values of $\beta$. In \eqref{mtildebetaz1}, these poles arise from the $\sin(\pi \beta/2)$ factor. The residues at these poles encode the Wilson coefficients in the low energy expansion of the amplitude in momentum space \cite{Arkani-Hamed:2020gyp}.

%%%%%%%%%%%%%%%%%%%%%%%%%%%%%%%%%%%%%%%%%%%%%%%%%%%%%%%%%%%%%%%

%\section{Low energy constraints}

\section{Celestial insight 2: Moment bounds}\label{sec4}

In section \ref{subsec:RMTapp} below, we use Ramanujan's master theorem to relate the Wilson coefficients in the low energy expansion of $2$-$2$ scalar amplitudes to the residues of the corresponding Mellin amplitude $\widetilde{\mathbf{M}}(\beta,z)$ at negative integer values of $\beta$. We then discuss locality constraints in the context of CSDR in section \ref{subsec:lc} and use it to obtain bounds on partial wave moments in section \ref{subsec:pwmomentbounds}. In section \ref{subsec:dnmpos} we  derive sufficient conditions for the positivity properties mentioned in section \ref{sec:newpositivity} to hold.      

\subsection{Applying Ramanujan's master theorem}
\label{subsec:RMTapp}

We now consider the low energy expansion of the amplitude $\mathbf{M}(s,t)$. If we do not include loop-level contribution of exchange of massless particles, then $\mathbf{M}(s,t)$ can be expanded around low energies as
\begin{equation}
\label{eftexp}
\begin{split}
\mathbf{M}(s,t) = \sum_{p,q=0}^{\infty} \mathcal{W}_{p,q} \hspace{0.04cm} x^{p}y^{q}
\end{split}
\end{equation}

where $x=-(st +tu +su), y=-s t u$ with $s+t+u=0$ and $\mathcal{W}_{p,q}$ denote Wilson coefficients. Let us now employ the change of variables $s=\omega^{2}, t=-z\omega^{2}, u=(z-1)\omega^{2}$. Then \eqref{eftexp} becomes
\begin{equation}
\label{eftexp1}
\begin{split}
\mathbf{M}(\omega^{2}, -\omega^{2} z) 
& = \sum_{n=0}^{\infty} \widetilde{\mathcal{W}}(n, z) \ \omega^{2n}
\end{split}
\end{equation}

where we have defined
\begin{equation}
\label{wzntilde}
\begin{split}
\widetilde{\mathcal{W}}(n, z) = \sum_{\substack{p,q \\ 2p+3q=n}} \mathcal{W}_{p,q} \hspace{0.05cm} z^{q}(z-1)^{q} (z^{2}-z+1)^{p}
\end{split}
\end{equation}

Now let us evaluate the Mellin amplitude $\widetilde{\mathbf{M}}(\beta,z)$ using the representation of the amplitude $\mathbf{M}(\omega^{2}, -\omega^{2} z)$  given by \eqref{eftexp1}. This can be done using Ramanujan's master theorem (RMT) for obtaining the Mellin transform of a function given its Taylor series expansion coefficients. See section \ref{sec:RMT} for further details of this theorem. 

Then according to RMT, the Mellin transform of \eqref{eftexp1} is given by
\begin{equation}
\label{mteftexp}
\begin{split}
\widetilde{\mathbf{M}}(\beta,z) & = \int_{0}^{\infty} d\omega \ \omega^{\beta-1}\ \mathbf{M}(\omega^{2}, -\omega^{2} z)  = \frac{\pi e^{-i\pi \beta/2} }{2 \sin\left(\frac{\pi \beta}{2}\right)} \hspace{0.07cm} \widetilde{\mathcal{W}}(-\beta/2,z)
\end{split}
\end{equation}

In writing the above expression, we have assumed that $\widetilde{\mathcal{W}}(n,z)$ can be analytically continued away from integer values of $n$. Now note that the $(1/\sin(\pi \beta/2))$ factor in \eqref{mteftexp} has poles when $\beta= -2n, n \in \mathbb{Z}_{\ge 0}$\footnote{$\widetilde{\mathbf{M}}(\beta,z)$ also has poles at $\beta =2n, n \in \mathbb{Z}_{\ge 0}$. In \cite{Arkani-Hamed:2020gyp} it was pointed out that the residues at these poles encode the coefficients in the high-energy expansion of the amplitude. However in this paper we will not consider this, since we are mainly interested in the low-energy  expansion of the amplitude. }. The residue at these poles is 
\begin{equation}
\label{betares}
\begin{split}
\mathrm{Res}_{\beta= -2 n} \left[ \widetilde{\mathbf{M}}(\beta, z) \right]  & =   \widetilde{\mathcal{W}}(n,z) =  \sum_{\substack{p,q \\ 2p+3q=n}} \mathcal{W}_{p,q} z^{q}(z-1)^{q} (z^{2}-z+1)^{p}
\end{split}
\end{equation}

Thus the residues of the Mellin amplitude at $\beta=-2n$ encode the Wilson coefficients appearing in the low energy expansion of the momentum space amplitude \cite{Arkani-Hamed:2020gyp}. 

\vskip 4pt
Now we can also write \eqref{betares} in terms of the ``crossing-symmetric'' variable\footnote{We refer to this as ``crossing-symmetric'' since it is invariant under $z\rightarrow 1-z$.} $\rho =-1-2z^2+2z$, introduced in section \ref{sec:newpositivity}. Then \eqref{betares} takes the form
\begin{equation}
\label{betaresrhovar}
    \begin{split}
\mathrm{Res}_{\beta= -2 n} \left[ \widetilde{\mathbf{M}}(\beta, \rho) \right]  &  = \widetilde{\mathcal{W}}(n,\rho)  = \sum_{\substack{p,q \\ 2p+3q=n}}\frac{(-1)^q}{2^{p+q}} \mathcal{W}_{p,q} 
(1+\rho)^{q} (1-\rho)^{p} \\
& \equiv\sum_{m=0}^{[\frac{n}{2}]}  d_m^{(n)}(1+\rho)^m(1- \rho)^{[\frac{n}{2}]-m}
\end{split}
\end{equation}

%%%%%%%%%%%%%%%%%%%%%%%%%%%%%%%%%%%%%%%%%%%%%%%%%%%%%%%%%%%%%%

\subsubsection*{Physical interpretation of the $\rho$ variable}

Noting that $z^2-z=- t u/s^2$, and $\cos\theta=1+2t/s$, where $\theta$ is the scattering angle, we can express $\rho$ 
\begin{equation}
    \rho= -1-2z^2+2z = -\frac{1+\cos^2\theta}{2}\,.
\end{equation}

Therefore, $\rho=-1$ corresponds to either $z=0$ or $z=1$, i.e.,  $\theta=0$ or $\theta=\pi$, while $\rho=1$ corresponds to the roots of $z^2-z+1=0$ which are  $z = (-1)^{1/3}, -(-1)^{2/3}$. Then clearly $\rho=1$ maps to unphysical (analytically continued) values of $\theta$.  

%The only singularities in the spin-$J$ Celestial partial wave for $\beta=-2n$ in eq.(\ref{mtildebetaz1}) occurs at $\rho=1$. These singularities are precisely the non-local singularities which need to cancel in a local theory. Further note that 
%\begin{equation}
   % f(z)=\frac{z(z-1)}{z^2-z+1}\rightarrow \frac{\rho+1}{\rho-1}\leq 0 \,, \qquad {\rm when} -1\leq \rho\leq 1\,,
%\end{equation}

%%%%%%%%%%%%%%%%%%%%%%%%%%%%%%%%%%%%%%%%%%%%%%%%%%%%%%%%%%%%%%%

\subsection{Locality Constraints}
\label{subsec:lc}

In a local EFT, the low energy expansion of the amplitude \eqref{eftexp} only contains positive powers of $x$. This implies that in \eqref{wzntilde} $\widetilde{\mathcal{W}}(n,z)$ should  be non-singular at $z = (-1)^{1/3}, -(-1)^{2/3}$, which are the roots of $z^{2}-z+1=0$. In terms of the $\rho$ variable these points corresponds to $\rho=1$ as mentioned in the previous subsection. Consequently such singularities are not allowed in the residues of the celestial amplitude at $\beta=-2n, n \in \mathbb{Z}^{+}$ for a local theory\footnote{The singularities at $z = (-1)^{1/3}, -(-1)^{2/3}$ lie outside the domain of physical kinematics where $z\in (0,1)$. }.  However the absence of these singularities is not manifest when the celestial amplitude is evaluated using the crossing symmetric dispersive representation of the momentum space amplitude. This is essentially the fact that the crossing symmetric dispersion relation makes crossing symmetry manifest at the expense of locality. Let us now see this a bit more explicitly as follows.  

\vskip 4pt
Taking the residue at $\beta= -2n, n \in \mathbb{Z}^{+}$ on the R.H.S. of \eqref{mtildebetaz1} we get
\begin{equation}
\label{betarescsdr}
\begin{split}
& \mathrm{Res}_{\beta= -2 n} \left[ \widetilde{\mathbf{M}}(\beta, z) \right] = \widetilde{\mathcal{W}}(n,z) \\
& =     (-1)^{n}  \sum_{J=0}^{\infty} (2J+1) \  \widetilde{\alpha}_{J}(n,\delta_{0})    \bigg[  (-1)^{n}   P_{J}\left( |1-2 z| \right) + z^{n} P_{J}\left( \left| \frac{z-2}{z}\right| \right)   +   (1-z)^{n} P_{J}\left( \left| \frac{z+1}{z-1}\right| \right)   \\
&  +   (z(1-z))^{n-J} (z^{2}-z+1)^{3-n}  \mathcal{Q}_{J}(-2n,z) \bigg]  
\end{split}
\end{equation}
where
\begin{equation}
\label{pwmoments}
\begin{split}
\widetilde{\alpha}_{J}(n, \delta_{0}) = 32\int_{\delta_{0}}^{\infty}  \frac{ds}{s^{n+1}} \ \alpha_{J}(s)
\end{split}
\end{equation}

The $(z^{2}-z+1)^{3-n}$ factor in the second line of \eqref{betarescsdr} is singular at $z = (-1)^{1/3}, -(-1)^{2/3}$ for $n\ge 4$. It is also worth noting that there are also apparent divergences at $z=0,1$ in \eqref{betarescsdr}. But it can be easily checked that the $z=0,1$ singularities cancel for any fixed $J$. 

\vskip 4pt
Thus for a local theory we need to impose on \eqref{betarescsdr} the constraint that the $z = (-1)^{1/3}, -(-1)^{2/3}$ singularities cancel upon performing the sum over spins $J$ in the partial wave expansion. In order to study the implications of these constraints, which will henceforth be referred  to as the locality or null constraints, it again turns out to be convenient to use the $\rho$ variable. Then it can be shown that \eqref{betarescsdr} takes the following form
\begin{equation}
\label{betarescsdrrho}
\begin{split}
& \mathrm{Res}_{\beta=-2n} \left[ \widetilde{\mathbf{M}}(\beta,\rho)\right] = \sum_{J=0}^{\infty}(2J+1) \widetilde{\alpha}_{J}(n, \delta_{0}) \bigg[ \sum_{k=1}^{n-3} \frac{c_{k}(n,J)}{(\rho-1)^{k}} + \mathcal{F}_{B}(n, J, \rho) \bigg]
%& = \sum_{J=0}^{\infty}(2J+1) \widetilde{\alpha}_{J}(n, \delta_{0}) \bigg[ \sum_{k=1}^{n-3} \frac{c_{k}(n,J)}{(\rho-1)^{k}} + \sum_{m=0}^{\left[ \frac{n}{2}\right]} \chi^{(n)}_{m}(J) \ (1+\rho)^{m}(1-\rho)^{\left[ \frac{n}{2}\right]-m}\bigg]
\end{split}
\end{equation}
where
\begin{equation}
\label{cknJdef}
\begin{split}
c_{k}(n,J) & = -  \frac{ 2^{J-3}}{(n-3-k)!}\  \lim_{\rho \to 1} \frac{d^{n-3-k}}{d\rho^{n-3-k}}\bigg[ (1+\rho)^{n-J}\mathcal{Q}_{J}(-2n,\rho)\bigg] 
\end{split}
\end{equation}
and $\mathcal{F}_{B}(n, J, \rho) $ is a polynomial in $\rho$ of degree $[n/2]$. We will refer to this as the Feynman block. See the Appendix, section \ref{sec:formulas} for their closed form expressions. In section \ref{feynmanblock} where we analyse the properties of Feynman blocks we will present some explicit examples of these blocks for few values of $n$. 

Now demanding that the singularities at $\rho=1$ cancel for a local theory, we get
\begin{equation}
\label{locconstr}
\begin{split}
 \sum_{J=2}^{\infty} (2J+1) \  c_{k}(n,J) \  \widetilde{\alpha}_{J}(n,\delta_{0})  =0 \quad \quad \forall k=1,\cdots, n-3
\end{split}
\end{equation}
where the sum above runs only over even spins $J$. Also note that this sum starts from $J=2$, since $c_{k}(n,J=0) =0$. In section \ref{subsec:pwmomentbounds} we will use the above locality constraint equations to derive analytic bounds on partial wave moments. In section \ref{subsec:dnmpos}, equation \eqref{locconstr} will also play a crucial role in analysing the novel positivity properties of low-energy expansion of the amplitude mentioned before. For this it is useful to relate the coefficients $d^{(n)}_{m}$, which are in turn related to the Wilson coefficients via \eqref{betaresrhovar}, to the partial wave moments $\widetilde{\alpha}_{J}(n,\delta_{0})$. In order to obtain this relation, we impose the locality constraints in \eqref{betarescsdrrho} and compare with \eqref{betaresrhovar}. This yields,
\begin{equation}
\label{dalphatrel}
\begin{split}
&  d^{(n)}_{m} =  \sum_{J=0}^{\infty}(2J+1)\     \chi^{(n)}_{m}(J) \ \widetilde{\alpha}_{J}(n, \delta_{0}) , \quad \quad m=0,1,\cdots, \left[\frac{n}{2}\right]
\end{split}
\end{equation}

Explicit expressions for the coefficients $\chi^{(n)}_{m}(J)$ can be obtained using the results given in Appendix \ref{sec:formulas}.

%%%%%%%%%%%%%%%%%%%%%%%%%%%%%%%%%%%%%%%%%%%%%%%%%%%%%%%%%%%%%%

\subsection{Bounds on partial wave moments}
\label{subsec:pwmomentbounds}

%{\bf AS: This section still looks a tad algebra heavy with not much explanations. SG, please try to fix this.}
In this section we show that the locality constraint equations \eqref{locconstr} can be used to derive lower bounds on the moments of partial wave coefficients. 
We first consider the case $n=4$ in \eqref{locconstr} which yields
\begin{equation}
\label{locn4}
\begin{split}
 \sum_{J=2}^{\infty} (2J+1)  \ c_{1}(4,J)  \ \widetilde{\alpha}_{J}(4,\delta_{0})=0 
\end{split}
\end{equation}

Using \eqref{cknJdef} it can be shown that $c_{1}(4,J)$ is given by
\begin{equation}
\label{cn4J}
\begin{split}
&  c_{1}(4,J) = J (J+1)(J^{2}+J-8)
\end{split}
\end{equation}

From \eqref{cn4J} it is clear that  $c_{1}(4,2) <0 $ and $ c_{1}(4,J) >0,  \ \forall J \ge 4$. Then let us write \eqref{locn4} as  
\begin{equation}
\label{locn41}
\begin{split}
60\hspace{0.05cm} \widetilde{\alpha}_{2}(4,\delta_{0})=  \sum_{J=4}^{\infty} J (J+1) (2J+1)(J^{2}+J-8)  \ \widetilde{\alpha}_{J}(4,\delta_{0})
\end{split}
\end{equation}

Now in a unitary theory, the partial waves are non-negative and this implies $\widetilde{\alpha}_{J}(n,\delta_{0}) \ge 0$. Therefore each term in the sum on the R.H.S. of \eqref{locn41} is a positive quantity. As a result we get for any $J \ge 4$
\begin{equation}
\label{alphatn4bound1}
\begin{split}
\frac{\widetilde{\alpha}_{2}(4,\delta_{0})}{\widetilde{\alpha}_{J}(4,\delta_{0})} > \frac{1}{60} \ J(J+1)(2J+1)(J^{2}+J-8), \quad J\ge 4 
\end{split}
\end{equation}

For example considering $J=4,6,8$, the above inequality implies
\begin{equation}
\label{pwn4boundsample}
\begin{split}
\frac{\widetilde{\alpha}_{2}(4,\delta_{0})}{\widetilde{\alpha}_{4}(4,\delta_{0})} > 36, \quad \frac{\widetilde{\alpha}_{2}(4,\delta_{0})}{\widetilde{\alpha}_{6}(4,\delta_{0})} > 309.4, \quad \frac{\widetilde{\alpha}_{2}(4,\delta_{0})}{\widetilde{\alpha}_{8}(4,\delta_{0})} > 1305.6%, \quad \frac{\widetilde{\alpha}_{2}(4,\delta_{0})}{\widetilde{\alpha}_{10}(4,\delta_{0})} > 3927
\end{split}
\end{equation}

As a comparison, we quote the values obtained from the dilaton amplitude in type II string theory:\footnote{In obtaining \eqref{typeiivals} we have performed the partial wave expansion in terms of Legendre polynomials  as in the case of the massless scalar amplitude in four spacetime dimensions in \eqref{absp}.}
\begin{equation}
\label{typeiivals}
    \begin{split}
\frac{\widetilde{\alpha}_{2}(4,\delta_{0})}{\widetilde{\alpha}_{4}(4,\delta_{0})} \approx 62, \quad \frac{\widetilde{\alpha}_{2}(4,\delta_{0})}{\widetilde{\alpha}_{6}(4,\delta_{0})} \approx 1258, \quad \frac{\widetilde{\alpha}_{2}(4,\delta_{0})}{\widetilde{\alpha}_{8}(4,\delta_{0})} \approx 13708.
\end{split}
\end{equation}

Evidently the type II dilaton amplitude satisfies the bounds obtained in \eqref{pwn4boundsample}. Similarly we can also obtain analytic bounds for partial wave moments with $n > 4$ using the locality constraints. We present below a sampling of the results for $n=5,6,7$. 

\subsubsection*{$n=5$ : }

%\begin{equation}
%\label{pwn5bound}
%\begin{split}
%\frac{\widetilde{\alpha}_{2}(5,\delta_{0})}{\widetilde{\alpha}_{J}(5,\delta_{0})} > \frac{1}{1080} \ J(J+1)(2J+1)[ J(J+1)(2J(J+1)-43)+150], \quad J\ge 4
%\end{split}
%\end{equation}

%This yields for example
\begin{equation}
\label{pwn5boundsample}
\begin{split}
\frac{\widetilde{\alpha}_{2}(5,\delta_{0})}{\widetilde{\alpha}_{4}(5,\delta_{0})} > 15, \quad \frac{\widetilde{\alpha}_{2}(5,\delta_{0})}{\widetilde{\alpha}_{6}(5,\delta_{0})} > 946.4, \quad \frac{\widetilde{\alpha}_{2}(5,\delta_{0})}{\widetilde{\alpha}_{8}(5,\delta_{0})} > 8411.6%, \quad \frac{\widetilde{\alpha}_{2}(4,\delta_{0})}{\widetilde{\alpha}_{10}(4,\delta_{0})} > 3927
\end{split}
\end{equation}

\subsubsection*{$n=6$ : }

%\begin{equation}
%\label{pwn6bound}
%\begin{split}
% & \widetilde{\alpha}_{2}(6,\delta_{0}) + 6\hspace{0.05cm}  \widetilde{\alpha}_{4}(6,\delta_{0}) \\
% & > \frac{1}{8640} (2J+1) (J-3) J (J+1) (J+4) \left(J (J+1) \left(J^2+J-32\right)+204\right) \widetilde{\alpha}_{J}(6,\delta_{0}), \quad J\ge 6
%\end{split}
%\end{equation}

%With $J=6,8$ for example in \eqref{pwn6bound} we get
\begin{equation}
\label{pwn6boundsample}
\begin{split}
& \frac{\widetilde{\alpha}_{2}(6,\delta_{0})}{\widetilde{\alpha}_{6}(6,\delta_{0})} +  \frac{6\hspace{0.05cm} \widetilde{\alpha}_{4}(6,\delta_{0})}{\widetilde{\alpha}_{6}(6,\delta_{0})}  > 1183, \quad \frac{\widetilde{\alpha}_{2}(6,\delta_{0})}{\widetilde{\alpha}_{8}(6,\delta_{0})} +  \frac{6\hspace{0.05cm} \widetilde{\alpha}_{4}(6,\delta_{0})}{\widetilde{\alpha}_{8}(6,\delta_{0})}  > 26214 
%& \widetilde{\alpha}_{2}(6,\delta_{0}) + 6\hspace{0.05cm}  \widetilde{\alpha}_{4}(6,\delta_{0})  > 230153 \hspace{0.05cm}  \widetilde{\alpha}_{10}(6,\delta_{0})
\end{split}
\end{equation}

Note that in the $n=6$ case, from \eqref{locconstr} we can only obtain bounds on the sum of ratios of partial wave coefficients as given in \eqref{pwn6boundsample}. This is because there happens to be only one independent locality constraint equation for $n=6$ and the coefficients $c_{1}(6, J=2), c_{1}(6,J=4)$ are both positive, while for $J\ge 4$ we have $c_{1}(6, J) <0$. 

\subsubsection*{$n=7$ : }

\begin{equation}
\label{pwn7boundsample1}
\begin{split}
& \frac{\widetilde{\alpha}_{2}(7,\delta_{0})}{\widetilde{\alpha}_{6}(7,\delta_{0})} > 200.2, \quad \frac{\widetilde{\alpha}_{2}(7,\delta_{0})}{\widetilde{\alpha}_{8}(7,\delta_{0})} > 38283.5, \quad  \frac{\widetilde{\alpha}_{4}(7,\delta_{0})}{\widetilde{\alpha}_{6}(7,\delta_{0})} > 30.33, \quad \frac{\widetilde{\alpha}_{4}(7,\delta_{0})}{\widetilde{\alpha}_{8}(7,\delta_{0})} > 338.38
\end{split}
\end{equation}

The inequalities in \eqref{pwn7boundsample1} can be derived by noting that for $n=7$, we get two independent null constraint equations from \eqref{locconstr}. For both these equations, we have $c_{k}(7,2), c_{k}(7,4) <0$ and $c_{k}(7,J) >0$ for $J\ge 6$. We have also checked that all the inequalities quoted above are satisfied by the type II string dilaton amplitude.

%\begin{equation}
%\label{pwn7bound1}
%\begin{split}
 % \widetilde{\alpha}_{2}(7,\delta_{0}) + 24.9\hspace{0.05cm}  \widetilde{\alpha}_{4}(7,\delta_{0}) &  > \frac{1}{43200} J (J+1) \bigg[J (J+1) (J (J+1) (J (J+1) (2 J (J+1)-155)+4836) \\
 % & \hspace{3.5cm} -65468)+235200 \bigg]  (2J+1) \widetilde{\alpha}_{J}(7,\delta_{0}), \quad J\ge 6
%\end{split}
%\end{equation}

%\begin{equation}
%\label{pwn7bound2}
%\begin{split}
 % \widetilde{\alpha}_{2}(7,\delta_{0}) + 27 \hspace{0.05cm}  %\widetilde{\alpha}_{4}(7,\delta_{0}) &  > \frac{1}{43200} J (J+1) \bigg[J (J+1) (J (J+1) (J (J+1) (2 J (J+1)-155)+4916) \\
  %& \hspace{3.5cm} -67908)+246960 \bigg] %(2J+1)\widetilde{\alpha}_{J}(7,\delta_{0}), \quad J\ge 6
%\end{split}
%\end{equation}

%With $J=6$ we get from \eqref{pwn7bound1}
The inequalities in eqs.\eqref{pwn4boundsample}-\eqref{pwn7boundsample1} demonstrate the phenomenon of {\it low spin dominance}(LSD) and we will say more about this in section (\ref{sec:lsd}).
%%%%%%%%%%%%%%%%%%%%%%%%%%%%%%%%%%%%%%%%%%%%%%%%%%%%%%%%%%%%%%%%%%%%%%
\subsection{Investigating $\Prho$}
\label{subsec:dnmpos}

In this section we further explore the $\mathcal{P}_{\rho}$ positivity property of the low energy expansion  of $2$-$2$ amplitude of massless scalars in four spacetime dimensions,  by considering the relation \eqref{dalphatrel} between the coefficients $d^{(n)}_{m}$ and the partial wave moments obtained using the crossing symmetric dispersion relation.

The above mentioned positivity property implies that we should have $d^{(n)}_{m} \ge 0$. For $n=2k, k\in \mathbb{Z}$ and $m=0$, this can be shown to follow from the fact that in a unitary theory the partial waves are positive. However in general,  unitarity alone does not imply $d^{(n)}_{m} \ge 0$ for any $n$. Here we argue that if the spin $J=0$ contribution to the partial wave decomposition of the amplitude dominates over the contribution from higher spins, then the positivity feature holds. We shall illustrate this below for $n=5$ and derive the sufficient condition for $d^{(5)}_{m} \ge 0$ to hold. Further examples for other values of $n$ are considered in the Appendix, section \ref{sec:dposexamples}. 

We begin by considering \eqref{dalphatrel} for $n=5$ which is given by
\begin{equation}
\label{dn5m}
\begin{split}
&  d^{(5)}_{m} = \sum_{J=0}^{\infty}(2J+1)\ \chi^{(5)}_{m}(J) \  \widetilde{\alpha}_{J}(5, \delta_{0}) , \quad \quad m=0,1,2
\end{split}
\end{equation}

For $m=0$ and $m=2$, it can be easily checked that the R.H.S. of \eqref{dalphatrel} is identical to the locality constraint equations for $n=5$. This immediately gives $d^{(5)}_{0} =d^{(5)}_{2} =0$. The only non-trivial case here is then $n=5,m=1$. In terms of the Wilson coefficients $\mathcal{W}_{p,q}$'s we have $d^{(5)}_{1} = -\mathcal{W}_{1,1}/4$.

\vskip 4pt
Now for $m=1$, it can be shown that $\chi^{(5)}_{m}(J)$ is given by
\begin{equation}
\label{chin5m1}
\begin{split}
 \chi^{(5)}_{1}(J) % & =  \frac{1}{144} J (J+1) (5 J (J+1) (2 J (J+1)-43)+822)-\frac{5}{4}  \\
& = \left( \frac{5}{4}- \frac{1}{2} J(J+1)  \right) -   \frac{5}{24} J (J+1) ( J (J+1) (2 J (J+1)-43)+150)
\end{split}
\end{equation}

Let us note that for $n=5$ the locality constraint equation takes the form
\begin{equation}
\label{lcn5}
\begin{split}
&  \sum_{J=2}^{\infty}(2J+1) \   J (J+1) (J (J+1) (2 J (J+1)-43)+150)  \ \widetilde{\alpha}_{J}(5 , \delta_{0})  =0 
\end{split}
\end{equation}

Then substituting \eqref{chin5m1} in \eqref{dn5m} with $m=1$ and using \eqref{lcn5} we get
\begin{equation}
\label{dn5m1a}
\begin{split}
&  d^{(5)}_{1} =  \frac{1}{4} \bigg[  5 \hspace{0.05cm} \widetilde{\alpha}_{0}(5, \delta_{0}) -  \sum_{J=2}^{\infty}(2J+1) \left(2 J(J+1) - 5 \right) \ \widetilde{\alpha}_{J}(5, \delta_{0})  \bigg] 
\end{split}
\end{equation}

{\it This readily implies that unless spin-0 is present, positivity in $\rho\in (-1,1)$ cannot hold.} 

\vskip 4pt
Now we can derive a sufficient condition for $d^{(5)}_{1} \ge 0$  to hold as follows. We use \eqref{lcn5} to eliminate $\widetilde{\alpha}_{4}(5,\delta_{0})$ from \eqref{dn5m1a}. This yields
\begin{equation}
\label{dn5m1e}
\begin{split}
  d^{(5)}_{1}& =  \frac{1}{4} \bigg[  5 \hspace{0.05cm} \widetilde{\alpha}_{0}(5, \delta_{0}) - 56 \hspace{0.05cm} \widetilde{\alpha}_{2}(5, \delta_{0}) \\
& +  \frac{1}{360} \sum_{J=6}^{\infty}(2J+1)  (J-4) (J+5) \left(14 J^4+28 J^3-7 J^2-21 J-90\right) \ \widetilde{\alpha}_{J}(5, \delta_{0})  \bigg] 
\end{split}
\end{equation}

Since all terms in the second line of \eqref{dn5m1e} are positive, we see that for $d^{(5)}_{1} \ge 0$ to hold, it suffices to have
\begin{equation}
\label{dn5m1f}
\begin{split}
    \widetilde{\alpha}_{0}(5, \delta_{0}) \ge  11.2 \hspace{0.05cm} \widetilde{\alpha}_{2}(5, \delta_{0}) 
\end{split}
\end{equation}

We can obtain similar inequalities for higher values of $n$ as well. For example, for $n=6,7,8$ we find
\begin{equation}
\label{spin0boundsn678}
\begin{split}
     \widetilde{\alpha}_{0}(6, \delta_{0}) \ge  25 \hspace{0.05cm} \widetilde{\alpha}_{2}(6, \delta_{0}), \quad  \widetilde{\alpha}_{0}(7, \delta_{0}) \ge   10.72 \hspace{0.05cm} \widetilde{\alpha}_{2}(7, \delta_{0})\,,\quad \widetilde{\alpha}_{0}(8, \delta_{0}) ~\ge~   13.75 \hspace{0.05cm} \widetilde{\alpha}_{2}(8, \delta_{0})\,.
     \end{split}
\end{equation}

%%%%%%%%%%%%%%%%%%%%%%%%%%%%%%%%%%%%%%%%%%%%%%%%%%%%%%%%%%%%%%%

%%%%%%%%%%%%%%%%%%%%%%%%%%%%%%%%%%%%%%%%%%%%%%%%%%%%%%%%%%%%%%%%%%%%%%%%%
\subsection*{How common is this positivity?}

{\bf In presence of spin 0}: Using the locality constraints we had already obtained conditions that suggested spin-2 dominance, for instance via equations \eqref{pwn4boundsample} and \eqref{pwn5boundsample}. In the positivity analysis above, we saw that if there is spin-0 dominance, then there is a novel positivity which was alluded to in section 1. 
These considerations enable us to make the following observations. The type II string tree level suggests that $\widetilde \alpha_2(n,\delta_0=1)\lesssim 0.04 \exp(-n/\sqrt{2})$. Consider for instance the $n=6$ case. This would give $\widetilde \alpha_0(6,1)\gtrapprox 0.01$. Now if $\widetilde \alpha_0(6,1)$ takes on values between $0-1$ (the string answer is approximately 1.001), then we conclude that for random values for $\widetilde \alpha_0(6,1)$ in this range, there is a $99\%$ possibility for us to find that $d_n^{(6)}\geq 0$. Thus the question becomes, what range of $\widetilde \alpha_0(6,\delta_0)$ is typical? 
The discussion above suggests that whenever there is spin-0 dominance of the form 
\begin{equation} \label{lsd}
   \frac{ \widetilde \alpha_0(n,\delta_0)}{\widetilde \alpha_2(n,\delta_0)}\gtrsim O(10)\,,
\end{equation}
we will obtain positivity for these class of theories. 

\noindent{\bf In absence of spin-0}: A counterexample to the $\Prho$ positivity is the following toy amplitude:
\begin{eqnarray} \label{1by}
M(s,t,u)= \frac{m^4}{(m^2-s)(m^2-t)(m^2-u)} -\frac{4}{3}~ \tanh^{-1}\left({\frac{1}{3}}\right)\left(\frac{1}{m^2-s}+\frac{1}{m^2-t}+\frac{1}{m^2-u}\right)
\end{eqnarray}
where the second term has been chosen to make the spin-$0$ partial wave contribution vanish. One can easily check that all the higher spin partial waves and all their moments are positive in this case. However, we know of no local Lagrangian description which could give rise to this amplitude. Further, this amplitude seems to necessarily indicate the existence of an infinite tower of massive higher spin particles all of which have the same mass $m^2$. Theories with an {\it accumulation point} in the spectrum such the one above seem to play a role in S-matrix bootstrap \footnote{An analogous example in the case of Polyakov bootstrap is the 2-d Ising Mellin amplitude which also exhibits similar behaviour due to the presence of twist-0.}, though its not clear if they can be ruled out by other considerations\cite{sch1,yutintriple}.

\noindent We can then look at the low energy expansion of the amplitude (setting $m=1$ for brevity)
\begin{eqnarray}
M(s,t,u)&=& -0.386294 + 0.0758038~ x + 0.0758038~ x^2 + 0.0758038 ~x^3 + 0.0758038~ x^4 \nonumber\\ &+& 0.0758038 ~x^5 + 0.0758038 ~x^6 + 0.386294 ~y + 
 0.310491~ x~ y \nonumber\\ &+& 0.234687 ~x^2 ~y + 0.158883~ x^3 ~y + 0.0830793~ x^4 ~y + 
 7 ~x^5 ~y 
\end{eqnarray}
and notice that $\mathcal{W}_{1,1},\mathcal{W}_{3,1},\mathcal{W}_{5,1} \ge 0$. When the spin-0 partial is absent, it can be shown generally using results of our previous section that 
\begin{equation}
    d_1^{(2k+1)}\le 0 ,
\end{equation}
which in terms of $\mathcal{W}_{pq}$ reads:
\begin{equation}
    \mathcal{W}_{2k+1,1}\ge 0\,, \quad \forall k=(0,1,2\cdots).
\end{equation}

%%%%%%%%%%%%%%%%%%%%%%%%%%%%%%%%%%%%%%%%%%%%%%%%%%%%%%%%%%%%%%%
\subsection{Comments about LSD}\label{sec:lsd}
The inequalities on the ratio of partial wave moments that we find in \eqref{alphatn4bound1} and other such inequalities in eqs.\eqref{pwn4boundsample}-\eqref{pwn7boundsample1} demonstrate  the phenomenon of {\it low spin dominance} (LSD) previously considered in the context of gravitational EFT's \cite{sasha,rs2,yutin3} for spins $J\ge 2$. 
\begin{equation}\label{degoflsd}
  \frac{\widetilde{\alpha}_{2}(n,\delta_{0})}{\widetilde{\alpha}_{J}(n,\delta_{0})} \ge \lambda  
\end{equation}

Since spin-$0$ does not directly enter the locality constraints \eqref{locconstr}, we cannot quantify  $  \frac{\widetilde{\alpha}_{0}(n,\delta_{0})}{\widetilde{\alpha}_{J}(n,\delta_{0})}$ directly using locality. However as we have argued in the previous section if there is spin-$0$ dominance of the form given in \eqref{lsd} namely $  \frac{\widetilde{\alpha}_{0}(n,\delta_{0})}{\widetilde{\alpha}_{2}(n,\delta_{0})} \gtrapprox 10$ then we can readily translate this to obtain:
\begin{equation}
\label{lsdbound}
 \frac{\widetilde{\alpha}_{0}(n,\delta_{0})}{\widetilde{\alpha}_{J}(n,\delta_{0})} \ge 10 \hspace{0.05cm}\lambda
\end{equation}

Since these follow directly from the locality constraints \eqref{locconstr} we can conclude that \\

{\it Low spin dominance (LSD) for $J\ge 2$ in scalar low energy EFT's is a consequence of locality.}\\

\noindent Furthermore, we can also quantify the parameter $\lambda$\footnote{ This parameter is usually been denoted by $\alpha$ in the  literature but to avoid confusion with the partial wave moments we denote it by $\lambda$ in this work}, which has been referred to as the {\it degree of LSD} in the literature \cite{sasha,yutin3,yutin4}. Our analysis  indicates that $\lambda$ in \eqref{lsdbound} is a function of $n$. For example,  we  find $\lambda \equiv \lambda(n)$, we have $\lambda(4)=62,\lambda(5)=15,\lambda(7)=6.6$. We leave a more complete analysis of the properties of $\lambda(n)$ for future work.

\section{Celestial insight 3: Typically Realness}\label{sec5}

\subsection{Feynman blocks in $\rho$}
\label{feynmanblock}

We shall now discuss the positivity properties of the Feynman block $\mathcal{F}_B(n,J,\rho)$ with respect to $\rho$ for fixed value of $\beta=-2~n$ with $n$ being a positive integer. The Feynman block is a polynomial in $\rho$ of degree $\lfloor \frac{n}{2} \rfloor$:
\begin{equation}\label{fbclosedform}
\mathcal{F}_B(n,J,\rho) = \sum_{i=0}^{\lfloor \frac{n}{2}\rfloor} c_n(J)\rho^{i}
\end{equation}

\noindent A closed form expression for the Feynman block $\mathcal{F}_B(n,J,\rho)$ is given by  \eqref{fbclosed} in appendix \ref{sec:formulas}. For the first few values of $n$ these are as follows:
\begin{eqnarray}
\mathcal{F}_B(1,J,\rho)&=&0 \nonumber\\
\mathcal{F}_B(2,J,\rho)&=& 1-\rho \nonumber\\
\mathcal{F}_B(3,J,\rho)&=& \frac{-1}{2}(2 J^2+2 J-3)(1+\rho) \nonumber\\
\mathcal{F}_B(4,J,\rho)&=& \frac{1}{4}((-3 J^4-6 J^3+21 J^2+24 J+2)+\left(-J^4-2 J^3+7 J^2+8 J-4\right) \rho +2 \rho^2) \nonumber\\
\mathcal{F}_B(5,J,\rho)&=& \frac{1}{72}\left((90 - 786 J - 571 J^2 + 420 J^3 + 185 J^4 - 30 J^5 - 10 J^6) \right.\nonumber\\
&+& \left.(-J (1 + J) (150 - 43 J - 41 J^2 + 4 J^3 + 2 J^4)) \rho +18 (-5 + 2 J + 2 J^2) \rho ^2 \right) \nonumber\\
\end{eqnarray}

One can verify that the tree-level type-II string amplitude answer can be expanded in these blocks and the convergence in spin is fast. Further, it can be readily checked that for sufficiently large values of $J$ and for any value $n \ge 3 $ the Feynman block $\mathcal{F}_B(n,J,\rho)$ is positive for real $\rho$ in the interval $-1\leq \rho\leq 1$. In particular for any value of $n\ge 3$ there exists a critical value  $J=J_c(n)$ such that for $J \le J_c(n)$ we have  $\mathcal{F}_B(n,J,\rho)<0 $ and for $J > J_c(n)$ we have $\mathcal{F}_B(n,J,\rho)>0 $. This can be seen from the grid plot below.

\begin{figure}[H]
\centering
  \includegraphics[width=\linewidth]{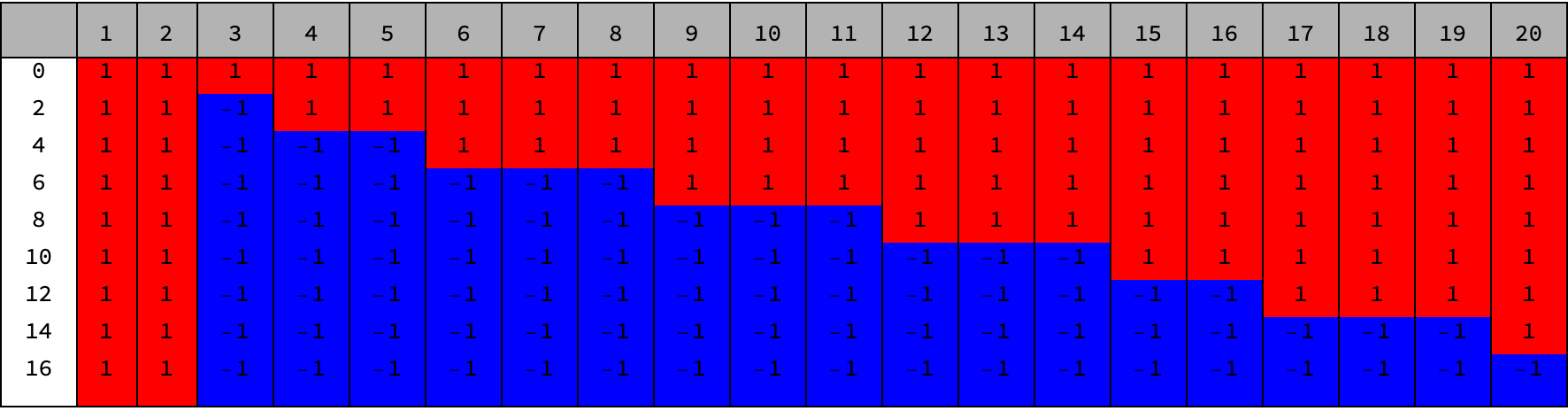}
  \caption{The above grid plots shows the signs of the Feynman block for $-1\le \rho\le 1$ for $0\le J \le 16$ and $1 \le n\le 20$. The red and blue indicate $J<J_c(n)$ and $J\ge J_c(n)$ corresponding to  negative and  positive signs respectively.}
  \label{fig:pic1}
\end{figure}

\subsection{GFT techniques}

We shall now discuss the connection of amplitude in the $\rho$ variable with typically realness. In \cite{rs1}, it was shown that the amplitude for appropriate range of the parameter\footnote{The parameter is given by $a=\frac{s t u}{s t+ tu +u s}$ where Mandelstam variables were parametrized via
$$
s=a \left(1-\frac{(\zeta-1)^3}{\zeta^3-1}\right),
t=a \left(1-\frac{(\zeta-e^{2\pi i/3})^3}{\zeta^3-1}\right)\,,
$$
Here $e^{2\pi i/3}$ is one of the cube-roots of unity. The $a$ works out to be 
$a=\frac{\omega'{}^2}{9}(\zeta^3-1)(1-\frac{1}{\zeta})^3\,$ which can curiously be related to the $z=-t/s$ variable using an $SL(2,\mathbb{C})$ transformation.} $a$ was a typically real function of $\zeta$ in the unit disk $|\zeta|<1$ and this connection proved quite fruitful for getting bounds on the Wilson coefficients. 

We shall briefly introduce the necessary background about typically real functions that we will need now and refer the interested reader to \cite{rs1,tatarczak} and references therein for further details. A typically real function $f(z)$ on a domain $\Omega \in \mathbb{C}$ which contains part of the real line $\mathbb{R}$ is defined as: 
\begin{equation}\label{TR}
    \Im f(z) \Im z >0 ~~{\rm for ~all}~~ z~~ {\rm such ~that}~~ \Im z \neq0\,.
\end{equation}
where $\Im f(z)$ means the imaginary part of the function $f(z)$.
It follows directly from the above definition that a typically real function $f(z)$ satisfies the following (see \cite{Wigner,goodman,robertson} and \cite{rs1} for a recent review):
\begin{enumerate}
    \item All poles of $f(z)$ lie on the real axis.
    \item All poles of $f(z)$ are simple.
    \item Residues at any pole of $f(z)$ is negative.
    \item Linear combinations of typically real functions with positive coefficients are typically real.
\end{enumerate}
Let us look at three physical examples of amplitudes that are typically real:
\begin{itemize}
    \item Consider the $\phi^2 \psi$ tree level amplitude of massless scalars $\phi$ with massive exchange $\psi$ \footnote{The $\alpha$,$\beta$,$\gamma$ and $\lambda$ used on this page are stand alone symbols not to be confused with symbols in other sections.} of mass $m$:
    \begin{eqnarray}
    M^{\phi^2 \psi}(\omega^2,\tilde{\rho})&=& \left(\frac{1}{s-m^2}+ \frac{1}{t-m^2}+\frac{1}{u-m^2}\right)\,, \nonumber\\
    &=& \frac{1}{m^2}\left(\frac{1}{(\lambda-1)}- \frac{2(2+\lambda)}{2+2 \lambda+\lambda^2 (1+\rho)} \right)\,,\nonumber\\
    \end{eqnarray}
where, $\lambda=\frac{\omega^2}{m^2}$. We see that the above is just a simple pole at $\rho=-1-\frac{2}{\lambda}(1+\frac{1}{\lambda})$  with negative residue $\lambda>0$. We can manually check its typically real as with $\rho=r e^{i \theta}$:
\begin{equation}
 \Im  M^{\phi^2 \psi}(\omega^2,\tilde{\rho})~ \Im \rho  = \frac{2 r \lambda^2 (\lambda+2) \sin^2 (\theta )}{r^2 \lambda^4 \sin ^2(\theta )+\left(r \lambda^2 (1+r \cos (\theta ))+2 \lambda+2\right)^2} >0 \,,
\end{equation}
for $\lambda>0$ and $\omega^2,m^2>0$.
Thus by the properties described above this is a typically real function for any value of $\omega^2$ except $\omega^2=m^2$ in an arbitrarily large disk around the origin in the $\rho$ variable. 
\item Consider next the type-II string amplitude eq.\eqref{type2} and in terms of the $\rho$ variable we can rewrite it using the infinte product representation of the gamma function as:
\begin{equation}
\mathcal{M}_{II} = \lambda_1(\omega^2) ~\frac{ \prod_{i=1}^{\infty} \left( 1+ \alpha_i ~\rho \right)} {\prod_{i=1}^{\infty} \left(1- \beta_i \rho \right)}\,
\end{equation}
where, $\alpha_n(\omega^2)=-\frac{\omega ^4 \left(n-\omega ^2\right)}{\left(n+\omega ^2\right) \left(2 n^2-2 n
   \omega ^2+\omega ^4\right)}$ , $\beta_n(\omega^2)=-\frac{\omega ^4 \left(n+\omega ^2-1\right)}{\left(-n+\omega ^2+1\right) \left(2 n^2+2 n
   \omega ^2-4 n+\omega ^4-2 \omega ^2+2\right)}$ and $\lambda_1(\omega^2) >0$. In fact, one can check numerically on Mathematica that for $\omega^2<2$ the amplitude is typically real directly by using \eqref{TR} for $|\rho|<1$! 
\item The closed bosonic string amplitude eq.\eqref{CBamplitude} can also be similarly written in terms of the $\rho$ variable as:
\begin{equation}
\mathcal{M}_{CB} = \lambda_2(\omega^2) ~\frac{ \prod_{i=1}^{\infty} \left( 1+ \gamma_i ~\rho \right)} {\prod_{i=1}^{\infty} \left(1- \delta_i \rho \right)}
\end{equation}
where, 
\begin{eqnarray}
\gamma_n(\omega^2)&=&\frac{9 \omega ^4 \left(-3 n+3 \omega ^2+1\right)}{\left(3 n+3 \omega ^2-1\right)
   \left(18 n^2-18 n \omega ^2-12 n+9 \omega ^4+6 \omega ^2+2\right)}\,, \nonumber\\ \delta_n(\omega^2)&=&-\frac{9 \omega ^4
   \left(3 n+3 \omega ^2-2\right)}{\left(-3 n+3 \omega ^2+2\right) \left(18 n^2+18 n
   \omega ^2-24 n+9 \omega ^4-12 \omega ^2+8\right)} \nonumber \,
   \end{eqnarray}
   and $\lambda_2(\omega^2) >0$. In this case as well one can check numerically on Mathematica that for large ranges of $\omega^2$ the amplitude is typically real directly by using  \eqref{TR} for $|\rho|<1$ ! 
\end{itemize}
The above motivates one to investigate positivity and typical realness  in $\rho$ more carefully.
To see the power of typical-realness let us look at a generic crossing symmetric monomial $x^p y^q= \frac{(-1)^q}{2^{p+q}}\omega^{4 p+ 6 q}(1-\rho)^p(1+\rho)^q$ that could appear in the low energy expansion of the amplitude. We could ask when such a term is typically real in the disk $|\rho|<r$ ? 

\noindent The possible terms fall into one of 4 categories $x^p y^q$  with either $p,q \ge 0$, or $p\ge 0, q\le 0$ or $q\ge0 ,p\le 0$ or $p,q \le 0$. The latter two cases correspond to having poles at $z=(-1)^{1/3},-(-1)^{2/3}$ and can be ruled out if we assume locality as we explained in section \ref{subsec:lc}. Analyzing the first two cases which we shall call (R)egular and (S)ingular respectively more carefully now by applying the definition \eqref{TR} we get the following possibilities listed in the table below:
\begin{center}
\begin{tabular}{| c | c | c| c|} 
\hline
Radius & Allowed & R & S\\
  \hline
   & & & \\
   &  &\(\displaystyle -(1-\rho )^n,-(\rho +1)^n,\) for &  \(\displaystyle -\frac{(1-\rho)^n}{(1+\rho)^m}, \)~ for \\
   & $$ \color{green} \CheckmarkBold $$& \(\displaystyle ~ 1\le n \le 2 \) & \(\displaystyle ~0\le n \le 2, 1\le m\le 2 \) \\
    $r=1$ & & &\\\cline{2-4}
       & & & \\
     &$$ \color{red} \XSolidBold $$& \(\displaystyle \pm(1-\rho)^n,\pm(1+\rho)^n,~ \forall n\ge 3\) &  \(\displaystyle -(1-\rho )^n,-(\rho +1)^n, \forall n \ge 3 \)\\
     & & & \\
  \hline
  & & & \\
   &  &\(\displaystyle -(1-\rho )^n,-(\rho +1)^n,\) for &  \(\displaystyle -\frac{(1-\rho)^n}{(1+\rho)^m}, \)~ for \\
   & $$ \color{green} \CheckmarkBold $$ & \(\displaystyle ~ 1\le n \le 2 \) & \(\displaystyle ~0\le n \le 1, m=1 \) \\
    $r=2$ & & &\\\cline{2-4}
       & & & \\
     &$$ \color{red} \XSolidBold $$& \(\displaystyle \pm(1-\rho)^n,\pm(1+\rho)^n,~ \forall n\ge 3\) &  \(\displaystyle -(1-\rho )^n,-(\rho +1)^n, \forall m \ge 2 \)\\
     & & & \\
  \hline
\end{tabular}
\end{center}

We can also shift to the $\tilde{\rho}=\rho+1$ variable as $\tilde{\rho}=0$ corresponds to either $t=0$ or $u=0$ which is a low energy limit. Thus in the new variable we can ask if the monomials are typically real in the disk $|\tilde{\rho}|<r$ and the possibilities are as below
\begin{center}
\begin{tabular}{| c | c | c| c|} 
\hline
Radius & Allowed & R & S\\
  \hline
   & & & \\
   &  &\(\displaystyle \tilde{\rho},~-(2-\tilde{\rho} )^n,\) for &  \(\displaystyle -\frac{(2-\tilde{\rho})^n}{\tilde{\rho}}, \)~ for \\
   & $$ \color{green} \CheckmarkBold $$& \(\displaystyle ~ 1\le n \le 6 \) & \(\displaystyle ~0\le n \le 3 \) \\
    $r=1$ & & &\\\cline{2-4}
       & & & \\
     &$$ \color{red} \XSolidBold $$& \(\displaystyle \tilde{\rho}^{n-4},~-(2-\tilde{\rho} )^{n+1},\forall n\ge 6\) &   \(\displaystyle -\frac{(2-\tilde{\rho})^n}{\tilde{\rho}^m}, ~\forall m \ge 2 \)\\
     & & & \\
  \hline
   & & & \\
   &  &\(\displaystyle \tilde{\rho},~-(2-\tilde{\rho} )^n,\) for &  \(\displaystyle -\frac{(2-\tilde{\rho})^n}{\tilde{\rho}}, \)~ for \\
   & $$ \color{green} \CheckmarkBold $$& \(\displaystyle ~ 1\le n \le 2 \) & \(\displaystyle ~0\le n \le 2 \) \\
    $r=2$ & & &\\\cline{2-4}
       & & & \\
     &$$ \color{red} \XSolidBold $$& \(\displaystyle \tilde{\rho}^{n},~-(2-\tilde{\rho} )^{n+1},\forall n\ge 2\) &   \(\displaystyle -\frac{(2-\tilde{\rho})^n}{\tilde{\rho}^m}, ~\forall m \ge 2 \)\\
     & & & \\
  \hline
\end{tabular}
\end{center}

Thus in disk $|{\tilde{\rho}}|<2$ there are precisely 6 possibilities that are allowed. Of these 3 of them are regular and correspond\footnote{The elements of class R usually come from the dispersive representation of the amplitude and are not expected to be individually typically real only the sum as a whole, so the above analysis does not rule out other regular terms.} to $y,x,x^2$. The three singular cases are $\frac{x^n}{y}$ for $0\le n \le 2$. The singular $n=0$ case is related to massless version of the tree level amplitude considered in \eqref{1by} and  singular cases corresponding to $n=1,2$ are respectively the exchange of massless scalar and spin-2 particles. Since the above cases are all typically real any sum of these with positive coefficients is typically real too. 
 
\subsection{Typically realness and $\widetilde{\mathbf{M}}(-2n,\rho)$}
%{\bf I think we should swap this to be earlier and make the graviton discussion after that.} \textcolor{red}{PR: I think its better if this subsection comes after the graviton since for getting the graviton bound we assume the whole amplitude is TR and we have a very different notion of TRness here for $J>J_T(n)$ and mixing things might be a little confusing for the reader. }
We shall now discuss the connection between typically realness and $\widetilde{\mathbf{M}}(\beta,\rho)$ at $\beta=-2n$. As discussed in \eqref{betaresrhovar}, \eqref{betarescsdrrho}.
\begin{eqnarray}\label{fbex}
    \mathrm{Res}_{\beta= -2 n} \left[ \widetilde{\mathbf{M}}(\beta, \rho) \right] &=& \sum_{J=0}^{\infty} (2J+1)~ \tilde{\alpha}_J(n,\delta_0) ~\mathcal{F}_B(n,J,\rho) \nonumber\\
    &=&\sum_{\substack{p,q \\ 2p+3q=n}}\frac{(-1)^q}{2^{p+q}} \mathcal{W}_{p,q} 
(1+\rho)^{q} (1-\rho)^{p}
\end{eqnarray}

The Feynman blocks $\mathcal{F}_B(n,J,\rho)$ are polynomials of degree $\left[ \frac{n}{2} \right]$ in $\rho$. By looking at $-\mathcal{F}_B(2,J,\rho)=\rho-1$ and $-\mathcal{F}_B(3,J,\rho)= \frac{1}{2}(2J^2+2J-3) (1+\rho)$ from the discussion in the previous section it is immediately obvious that these are typically real for any $J$ and $J\ge 2$ respectively. Thus, it is a natural question to ask if $\mathcal{F}_B(n,J,\rho)$ (possibly upto an overall sign) are typically real for any other values of $n,J$ inside the disk $|\rho|<1$? Since we have closed form expressions of the Feynman block $\mathcal{F}_B(n,J,\rho)$ in  eq. \eqref{fbclosedform} namely \eqref{fbclosed} for any $n,J$ this can be readily checked to sufficiently high values of $n,J$.
 Rather remarkably we find that the answer to the above question is in the affirmative and the result is as follows: \\
 
\noindent {\it {\emph The Feynman block $\mathcal{F}_B(n,J,\rho)$ (with appropriate sign) is a typically real polynomial of  degree $\lfloor\frac{n}{2} \rfloor$ in $\rho$ inside the unit disk $|\rho|<1$ for any value of $J\ge J_T$ with $J_T=n+2+\frac{1-(-1)^n}{2}$} for $n\ge 10$.}\\
 
\noindent For lower $n$, $J_T(n)$ can be read off from the table below:\\\\
\begin{tabular}{|c|c|c|c|c|c|c|c|c|c|c|c|c|c|c|c|c|c|}
\hline
 $n$ & 4& 5& 6& 7&8&9&10&11&12&13&14&15&19&20&21&99&100 \\
\hline
$J_T(n)$ &4 & 4 & 6 & 8& 8& 10 & 12& 14& 14&16 & 16 & 18 & 22 &22& 24&102&102\\
\hline
% etc. ...
\end{tabular} \\

\noindent In fact a well known family of typically real polynomials are called {\it Suffridge polynomials} \cite{suffridge1,suffridge2,suffridge3,shaffer1}   $S_{N,j}(\rho)$ and in all cases we have checked, $(-1)^n \mathcal{F}_B(n,J,\rho)$ for $n>3,J>J_T$ is a positive linear combination of these.
The Suffridge polynomials are defined as:
\begin{eqnarray}
S_{N,j}(\rho)&=&\sum_{k=1}^N A_k(N,j) \rho^k, ~~~{\rm with} \nonumber \\
A_k(N,j)&=& \frac{N-k+1}{N} \frac{\sin \left(\frac{k j \pi}{N+1}\right)}{\sin \left(\frac{ j \pi}{N+1}\right)}
\end{eqnarray}
A few examples are as follows:
\begin{eqnarray}
S_{2,4}&=&\rho -\frac{\rho ^2}{2} \nonumber\\
S_{3,1}&=&\frac{\rho ^3}{3}+\frac{2 \sqrt{2} \rho ^2}{3}+\rho \nonumber\\
S_{5,2}&=& -\frac{\rho ^5}{5}-\frac{2 \rho ^4}{5}+\frac{4 \rho ^2}{5}+\rho \nonumber\\
S_{7,3}&=&\frac{\rho^7}{7}+\frac{3}{7} \rho ^5 \cot \left(\frac{\pi }{8}\right)+\frac{5}{7} \rho ^3 \cot \left(\frac{\pi }{8}\right)-\frac{1}{7} \sqrt{2} \rho ^6 \csc \left(\frac{\pi
   }{8}\right)-\frac{4}{7} \rho ^4 \csc \left(\frac{\pi }{8}\right)-\frac{3}{7} \sqrt{2}
   \rho ^2 \csc \left(\frac{\pi }{8}\right)+\rho \nonumber
\end{eqnarray}
Some examples of Feynman blocks for $J \ge J_T$ as linear combinations of Suffridge polynomials with positive coefficients:
\begin{eqnarray}
\mathcal{F}_B(3,10,\rho)&=&\frac{217}{2}S_{1,1}+\frac{217}{2} S_{2,2}\nonumber\\
\mathcal{F}_B(4,6,\rho)&=&\frac{359}{2}S_{2,1}+\frac{357}{2} S_{2,2}
\end{eqnarray}
%{\bf AS: Convex hull of Suffridge polynomials?}
We quote the following results about typically real polynomials that will be useful for our purposes. The reader may refer to \cite{suffridge1,suffridge2,suffridge3,shaffer1,brandt} for details.  Let $f(z)=z+\sum_{i=2}^N a_i z^i$ be a typically-real polynomial of degree $N$ in the disk $|z|<1$ which we denote as $f(z) \in T^N$ and if $R(\cos{\theta}) =\frac{\Im f(e^{i \theta})}{\sin {\theta} }$. Then the following are true: 
\begin{enumerate}
    \item $f(z)\in T^N$ if and only if $R(\cos{\theta})= 1+\sum_{j=2}^N b_j \frac{\sin{j \theta}}{\sin{\theta}}\ge 0$.
    \item Let, $b_j \in \mathbb{R}$ with $b_{N-1}=1$ and if $\sum_{i=0}^{N-1} b_j u^j$ has a fixed sign for all $-1 \le u \le 1$ then $\exists$ unique $a_j$ with $1\le j\le N$ and $a_1=1$ such that $$R(u=\cos{\theta})=2^{N-1}a_N \sum_{j=0}^{N-1} b_j u^j$$ and $f(z)=z+\sum_{i=2}^{N } a_i z^i \in T^N$.
    \item If $f(z) \in T^N$ with $a_N \neq 0$ and for $1\le k\le N$, $a_k$ can assumes an extreme (max/min) value then all the zeros of $R(u)$ with $u=\cos{\theta}$ are real and 
\begin{subequations}\label{suffridgebounds}
  \begin{align}
  R(u)= 2^{N-1} a_N 
\begin{dcases*} 
\prod_{j=1}^{\frac{N-1}{2}}(u-\gamma_j)^2 
   & if  N is odd, $a_N > 0$\,, \\
(u^2-1)\prod_{j=1}^{\frac{N-3}{2}}(u-\gamma_j)^2 
   & if  N is odd, $a_N < 0$\,, \\
  (u+1) \prod_{j=1}^{\frac{N-2}{2}}(u-\gamma_j)^2 
   & if  N is even, $a_N > 0$\,, \\
   (u-1) \prod_{j=1}^{\frac{N-2}{2}}(u-\gamma_j)^2  
   & if  N is even, $a_N < 0$\,, 
\end{dcases*}
\end{align}
\end{subequations}
where, $-1\le \gamma_j\le 1$. 

\noindent In other words the coefficient body $\{(a_2,\cdots,a_N): z+a_2 z^2+\cdots+a_N z^N\}$ has extreme points that live on a manifold of dimension $\frac{N-1}{2},\frac{N-2}{2}$ for odd, even $N$ when $a_N>0$ and manifold of dimension $\frac{N-3}{2},\frac{N-2}{2}$ for odd, even $N$ when $a_N<0$ respectively.\\
Once one has the extreme points then the allow coefficinets region is a convex hull of these extreme points since the set of typically real polynomials is a convex set and Krein-Milman theorem \cite{rs1} applies. 
We work out the first couple of cases for illustrative purposes:

 {\bf N=2:} From eq.\eqref{suffridgebounds} assuming $a_2>0,a_2<0$ we get $ 2 a_2+2 a_2 u = 1+ 2 a_2 u, -2 a_2+2 a_2 u = 1+ 2 a_2 u$ respectively, which can be solved to get the extreme points $a_2=\pm\frac{1}{2}$ and convex hull of the extreme points yields the line $|a_2|\le \frac{1}{2}$.
 
 {\bf N=3:} From eq.\eqref{suffridgebounds} assuming $a_3>0,a_3<0$ we get $ 1-a_3+2 a_2 u+ 4 a_3 u^2 = 4 a_3 \gamma_1^2- 8 a_3 u \gamma_1 +4 a_3 \gamma_1^2, 1-a_3+2 a_2 u+ 4 a_3 u^2 = -4 a_3+4 a_3 \gamma_1^2$ respectively for $|\gamma_1|\le 1$, which can be solved to obtain $(a_2,a_3) =\left(\frac{-4 \gamma_1}{1+4 \gamma_1^2},\frac{1}{1+4 \gamma_1^2} \right)$ and $(a_2,a_3) =(0,-\frac{1}{3})$ respectively. Thus for $a_3>0$ we get part of the ellipse $(2a_3-1)^2+a_2^2=1$ with $a_3\ge \frac{1}{5}$ and for $a_3<0$ we get a point $\left(0,-\frac{1}{3}\right)$. The convex hull of which yields the 2-d region below.\\
 
 The regions are shown in the figure below along with some physical theories being marked as points. In the figure below green, blue,black  points denote tree level $\phi^2 \psi$ amplitude,  1-loop box $\phi^2 \psi$ amplitude and type-II string amplitude respectively. In left figure the grey point also denotes type-II string amplitude. since for both $n=4,5$ we get $N=2$ as the Feynman block is a degree $\lfloor\frac{n}{2} \rfloor$ polynomial we have denoted both in the figure. 
 \begin{figure}[H]
\centering
  \includegraphics[width=0.9\textwidth]{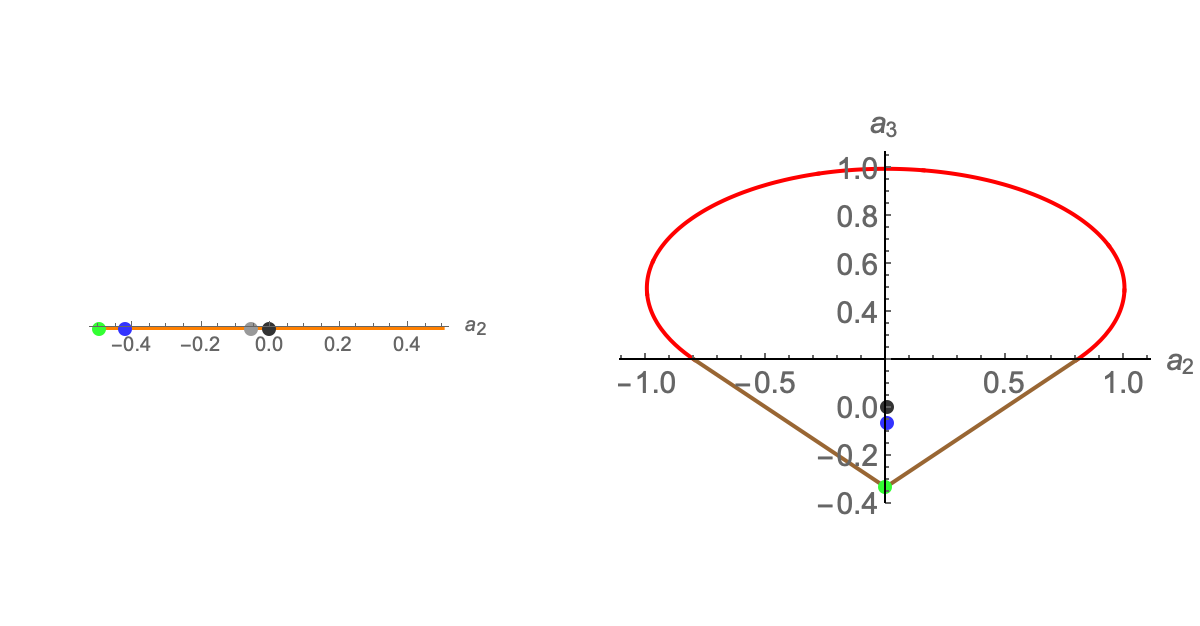}
  \caption{(left) The coefficient region for $N=2$ in the $a_2$ space.(Right)The coefficient region for $N=3$ in $(a_2,a_3)$ space.}
  \label{fig:pic4}
\end{figure} 
The red curve and the green points in the right side figure are the extremal manifolds and the brown lines have been added as the region is a convex hull of these extremal manifolds. In particular, the red curve gives us a one parameter family of extremal theories though none of the examples we are aware of seems to live on the red curve, it will interesting to find a theory that lives on the red curve. We leave this investigation for future work. However, tree level $\phi^2 \psi$ is an extremal theory that lives on the left most boundary and cusp of the $N=2,3$ cases respectively and this consistent with the observation that cubic tree level vertices are extremal in \cite{rs1}.

\noindent As is obvious this procedure will always give us a finite region which has implications for the $\mathcal{W}_{p,q}$ bounds we get, namely all $\mathcal{W}_{p,q}$'s will be two sided bounded as was shown using different methods in \cite{rs1}. 

\item For any $z$ inside the disk \cite{michel} we have 
\begin{equation}\label{suffridgedistortion}
|f(z)| \le \frac{1}{4}\csc^2{\frac{\pi}{2(N+2)}}\,.
\end{equation}
\end{enumerate}
The following are some plots of both Suffridge polynomials and the Feynman blocks:
\begin{figure}[H]
\centering
\begin{subfigure}{.3\textwidth}
    \centering
    \includegraphics[width=0.9\textwidth]{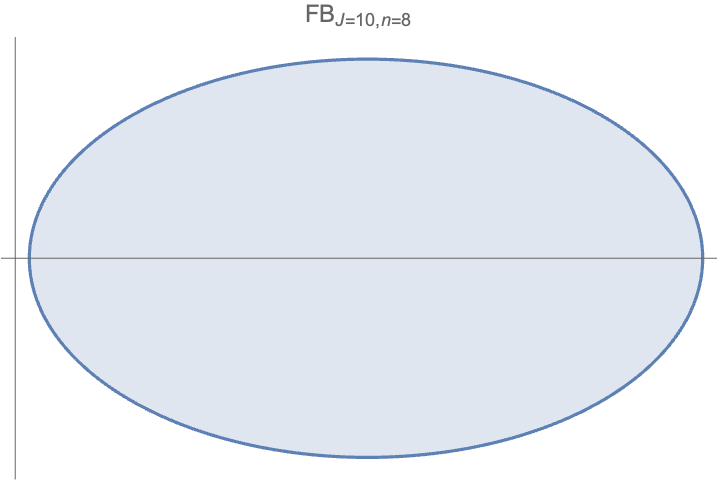}
\end{subfigure}%
\begin{subfigure}{.3\textwidth}
    \centering
    \includegraphics[width=0.9\textwidth]{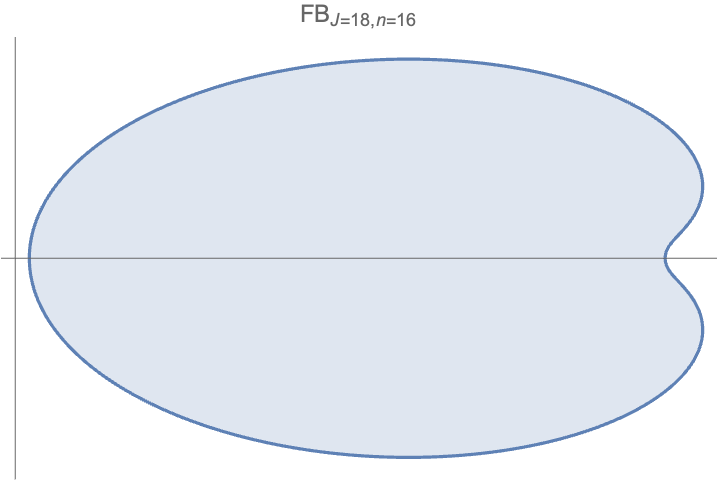}
\end{subfigure}
\begin{subfigure}{.3\textwidth}
    \centering
    \includegraphics[width=0.9\textwidth]{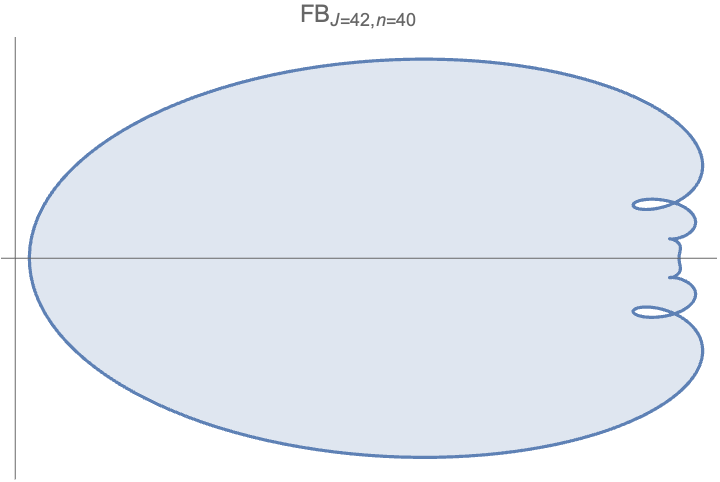}
\end{subfigure}%

\begin{subfigure}{.3\textwidth}
    \centering
    \includegraphics[width=0.9\textwidth]{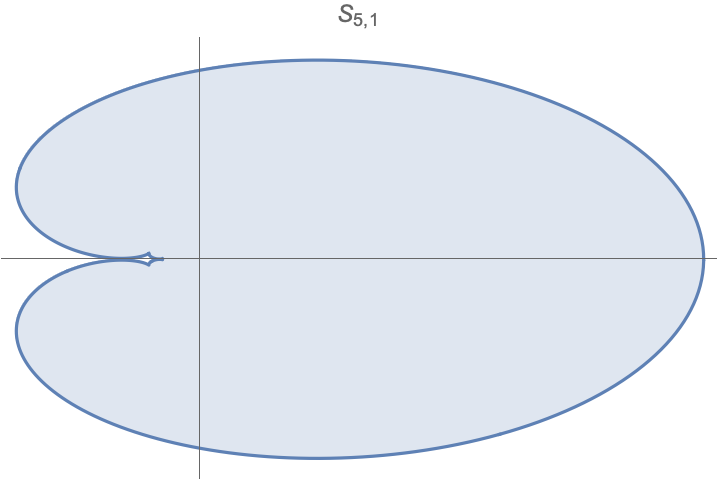}
\end{subfigure}%
\begin{subfigure}{.3\textwidth}
    \centering
    \includegraphics[width=0.9\textwidth]{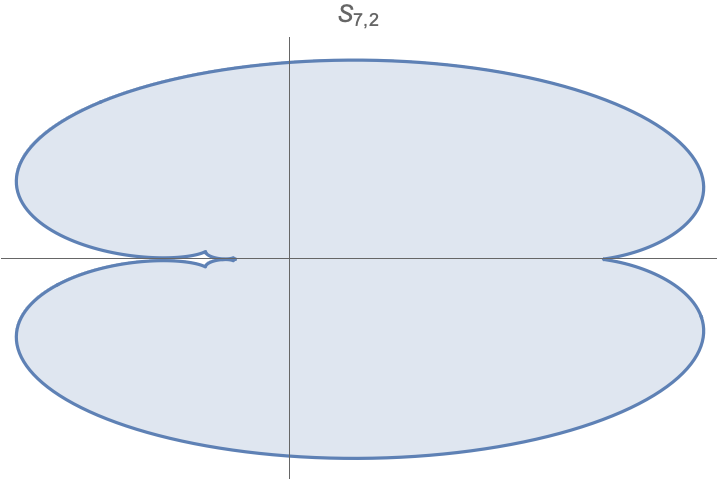}
\end{subfigure}
\begin{subfigure}{.3\textwidth}
    \centering
    \includegraphics[width=0.9\textwidth]{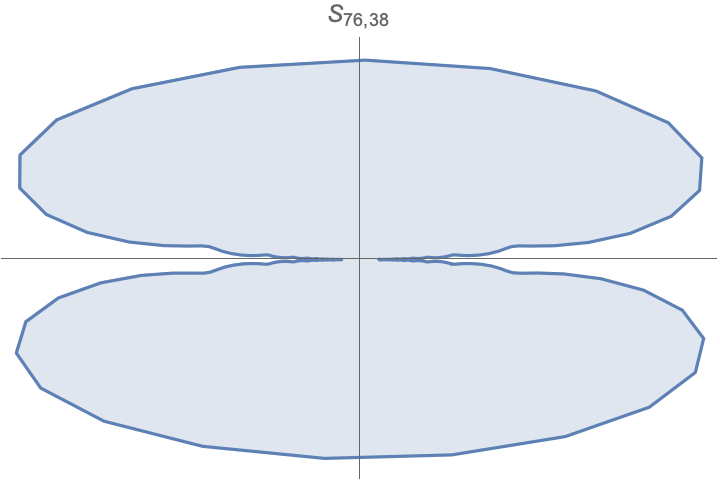}
\end{subfigure}%
\caption[short]{Some plots of Feynman blocks for different $n,J$ and Suffridge polynomials for different $N,j$.}
\end{figure}

\noindent The above discussion motivates us to consider the following:
\begin{eqnarray}
 \mathrm{Res}_{\beta= -2 n} \left[ \widetilde{\mathbf{M}}(\beta, z) \right] &=& \sum_{J=0}^{J_T-2} (2J+1)~ \tilde{\alpha}_J(n,\delta_0) ~\mathcal{F}_B(n,J,\rho)+\underbrace{\sum_{J=J_T}^{\infty} (2J+1)~ \tilde{\alpha}_J(n,\delta_0) ~\mathcal{F}_B(n,J,\rho)}_{R_T(n,\rho)} \nonumber
\end{eqnarray}
The $R_T(n,\rho)$ above being a positive sum of typically real functions is typically real.
Thus we have 
\begin{eqnarray} \label{Rtn1}
R_T(n,\rho) &=& \mathrm{Res}_{\beta= -2 n} \left[ \widetilde{\mathbf{M}}(\beta, z) \right] -\sum_{J=0}^{J_T-2} (2J+1)~ \tilde{\alpha}_J(n,\delta_0) ~\mathcal{F}_B(n,J,\rho)
\end{eqnarray}

\noindent An important observation is the following. Since the Feynman block $\mathcal{F}_B(n,J,\rho)$ for a particular $n$ is always a polynomial of the same degree $\lfloor\frac{n}{2}\rfloor$ for all $J$ so both the LHS and RHS of the equation \eqref{Rtn1} are polynomials of the same degree in $\rho$. This combined with the fact that $R_T(n,\rho)$ is a typically real polynomial allows one to get three kinds of constraints:
\begin{enumerate}
    \item {\bf Sign patterns $\mathcal{S}(n)$:} These follow since the Feynman blocks have a fixed sign pattern as we had discussed earlier in sec.(\ref{feynmanblock}). The signs of the coefficient of $\rho^k$ in $\mathcal{F}_B(n,J,\rho)$ is the same for all $J\ge J_T(n)$ and the sign of each term in $R_T(n,\rho)$ (being a positive sum of  $\mathcal{F}_B(n,J,\rho)$ for $J\ge J_T(n)$) is the same as that of $\mathcal{F}_B(n,J_T,\rho)$. Comparing this with RHS of \eqref{Rtn1} we get constraints on the coefficients. We shall denote these constraints by $\mathcal{S}(n)$.
    \item{\bf Suffridge bounds $TR(n)$:} These follow as $\tilde{R}_T(n,\rho)=\frac{R_T(n,\rho)-R_T(n,0)}{\partial_{\rho}R_T(n,\rho)|_{\rho=0}}$ is typically real polynomial and thus obeys the coefficient bounds arising from \eqref{suffridgebounds}. These depend on $n$ and can be worked out case by case for each $n$ as we had done for $n=2,3$ above. We list the first few cases in the table below:
    
{\scriptsize{\begin{tabular}{|c|c|c|c|c|c|}
\hline
   2& 3& 4& 5 & $N$ odd & $N$ even \\ 
\hline
  & $|a_2|\le1,$ & $|a_2|\le\frac{1+\sqrt{7}}{3},$& $|a_2|\le \sqrt{2},$ & $|a_2|\le 2\cos{\frac{2\pi}{N+3}}$ & $|a_2|\le 2\cos{\alpha}$ \\
  $|a_2|\le \frac{1}{2}$  & $ -1/3 \le a_3\le1$ & $ -1/3 \le a_3\le1$ & $ -\frac{\sqrt{5}-1}{2} \le a_3\le \frac{1-\sqrt{5}}{2}$ & \vdots  &  \vdots \\
   & & $ |a_4| \le 2/3$ & $ |a_4| \le 1$ & $|a_{N-1}|\le 1$  & $-\frac{N-2}{N+2}\le a_{N-1}\le 1$ \\
 &  &  & $ -\frac{1}{2}\le a_5\le 1$ & $-\frac{N-1}{N+3}\le|a_{N}|\le 1$ & $-\frac{N}{N+2}\le|a_{N}|\le \frac{N}{N+2}$ \\
\hline

% etc. ...
\end{tabular} }} \\

where, $\alpha \in \left(\frac{2 \pi}{N+4},\frac{2\pi}{N+2} \right)$ satisfies $(N+4)\sin{\frac{N+2}{2} \alpha}+(N+2)\sin{\frac{N+4}{2} \alpha}$ with $N=\lfloor n/2 \rfloor$.
   \item {\bf Distortion constraints $\mathcal{D}(n)$:} These also follow from \eqref{suffridgedistortion}. 
    \begin{eqnarray}
 |\tilde{R}_T(n,\rho)|   & \le & \frac{1}{4} \csc^2{\frac{\pi}{2(N+2)}},~~ {\rm with}~~N=\lfloor n/2 \rfloor
    \end{eqnarray}
The $\mathcal{S}(n)$,~$TR(n)$,~$\mathcal{D}(n)$ are respectively the analogues of polynomial analogues of the $PB_c,TR_U$ and distortion constraints used in \cite{rs1} for typically-real functions.   
\end{enumerate}
Let us consider the case $n=4$ which has $J_T=4$ in this case we get:
\begin{eqnarray}
R_T(4,\rho)&=& \frac{\mathcal{W}_{2,0}(1-\rho)^2}{4} -\sum_{0}^{2}(2J+1)\tilde{\alpha}_J(4) \mathcal{F}_B(4,J,\rho) \nonumber\\ &=&
-\frac{1}{4} (2 \tilde{\alpha}_0(4)+190 \tilde{\alpha}_2(4)-\mathcal{W}_{2,0})+\frac{1}{2}\left(2\tilde{\alpha}_0(4)-20\tilde{\alpha}_2(4)-\mathcal{W}_{2,0} \right)\rho \nonumber\\&& - \frac{1}{4}\left(2\tilde{\alpha}_0(4)+10 \tilde{\alpha}_2(4)-\mathcal{W}_{2,0} \right) \rho^2 \,, \\
\tilde{R}_T(4,\rho)&=& \rho -\frac{\left(2\tilde{\alpha}_0(4)+10 \tilde{\alpha}_2(4)-\mathcal{W}_{2,0}\right)}{2 \left(2\tilde{\alpha}_0(4)-20\tilde{\alpha}_2(4)-\mathcal{W}_{2,0} \right)}\rho^2 
\end{eqnarray}
We get the following: 
\begin{eqnarray}
\mathcal{S}(4)&:&~~~~~~~ 2 \tilde{\alpha}_0(4)+10 \tilde{\alpha}_2(4)\le \mathcal{W}_{2,0} \le 2 \tilde{\alpha}_0(4)+190 \tilde{\alpha}_2(4)\\  TR(4)&:&~~~~~~~ \mathcal{W}_{2,0} \ge 2 \tilde{\alpha}_0(4)-5 \tilde{\alpha}_2(4)  \\ \mathcal{D}(4)&:&  ~~~~~~~
|\tilde{R}_T(4,\rho)|\le 1+\frac{1}{\sqrt{2}}\approx 1.707 \\
&&~~~~~~~ \mathcal{W}_{2,0} \ge 2 \tilde{\alpha}_0(4)-20 \tilde{\alpha}_2(4)
\end{eqnarray}
Combining all of the above we get:
\begin{equation}\label{n4bounds}
2 \tilde{\alpha}_0(4)+10 \tilde{\alpha}_2(4) \le  \mathcal{W}_{2,0}  \le 2 \tilde{\alpha}_0(4)+190 \tilde{\alpha}_2(4)
\end{equation}
For the string case from appendix(\ref{sec:stringexamples}) this means $2.0164 \le \mathcal{W}_{2,0}\le 2.214$ and the string value is $ \mathcal{W}_{2,0}=2.016$ which is very close to the lower limit.
Using $\tilde{\alpha}_J(4) \le \frac{1}{4 M^8}$ and assuming $\frac{\tilde{\alpha}_0(4)}{\tilde{\alpha}_2(4)}>10$ we get the non-projective bound:
\begin{equation}
    \frac{1.5}{ M^8}\le \mathcal{W}_{2,0} \le \frac{10.5}{M^8}
\end{equation}A novel feature of the above is the existence of a lower bound $>0$.

Let us now consider the $n=5$ case with $J_T=4$  and 
\begin{eqnarray}
R_T(5)&=& - \mathcal{W}_{1,1} \frac{(1-\rho^2)}{4}-\sum_{J=0}^{2}(2J+1)\tilde{\alpha}_J(5) \mathcal{F}_B(5,J,\rho) \nonumber\\
&=&-\frac{1}{4} \left(5\tilde{\alpha}_0(5) +265 \tilde{\alpha}_2(5)+\mathcal{W}_{1,1} \right) -15 \tilde{\alpha}_2(5) \rho + \frac{1}{4}\left(5\tilde{\alpha}_0(5)-35\tilde{\alpha}_2(5)+\mathcal{W}_{1,1} \right) \rho^2 \,, \nonumber \\
 \tilde{R}_T(5,\rho)&=& \rho- \frac{\mathcal{W}_{1,1}+5    \tilde{\alpha}_0(5)-35\tilde{\alpha}_2(5)}{60\tilde{\alpha}_2(5) } \rho^2 \,.
\end{eqnarray}
We find
\begin{eqnarray}
~~\mathcal{S}(5)&:&~~~~~~~ \mathcal{W}_{1,1} \ge-5 \tilde{\alpha}_0(5)+35 \tilde{\alpha}_2(5) \\  TR(5)&:&~~~~~~~ -5 \tilde{\alpha}_0(5)+5 \tilde{\alpha}_2(5) \le \mathcal{W}_{1,1} \le -5\tilde{\alpha}_0(5)+65 \tilde{\alpha}_2(5) \\ \mathcal{D}(5)&:&~~~~~~~|\tilde{R}_T(5,\rho)|\le 1+\frac{1}{\sqrt{2}}\approx 1.707
\end{eqnarray}
Combining all of the above we get 
\begin{equation}\label{pw51}
-5 \tilde{\alpha}_0(5)+35 \tilde{\alpha}_2(5)   \le  \mathcal{W}_{1,1} \le -5 \tilde{\alpha}_0(5)+65 \tilde{\alpha}_2(5)\,.
\end{equation}
Using locality constraints which leads to \eqref{dn5m1e}, we get the following, slightly stronger, upper bound 
\begin{equation}\label{pw52}
 \mathcal{W}_{1,1} \le -5 \tilde{\alpha}_0(5)+ 56 \tilde{\alpha}_2(5)\,.
\end{equation}

The type II-string values for this case are $$\tilde{\alpha}_0(5)=1.0013,\tilde{\alpha}_2(5)=0.000538, \mathcal{W}_{1,1}=-4.9857 $$
In comparison our bound \eqref{pw51},\eqref{pw52} gives \begin{equation}-4.9878\le\mathcal{W}_{1,1} \le-4.9765 \end{equation} which is very narrow range and respected! Notice that this is a bound on the Wilson coefficient itself and not on a ratio. Using the fact that $\alpha_J(s)\leq 1$ and using 32$\int_{2M^2}^\infty \frac{ds}{s^{6}}\alpha_J(s)\leq \frac{1}{5 M^{10}}$, we have the non-projective bound
\begin{equation}
-\frac{1}{M^{10}}\leq \mathcal{W}_{1,1}\leq \frac{56}{5M^{10}}\,.
\end{equation}
Note that the type-II tree level $\alpha_J(s)$'s are to be thought of as distributions (since they have delta function support only) which leads to the different constraint given above.
%{\bf AS: Please quote the string values at each instance.}
\noindent Similarly, the $n=6$ case with $J_T=6$  and 
\begin{eqnarray}
R_T(6)&=& \mathcal{W}_{3,0} \frac{(1-\rho)^3}{8} + \mathcal{W}_{0,2} \frac{(1+\rho)^2}{4}-\sum_{J=0}^{4}(2J+1)\tilde{\alpha}_J(6) \mathcal{F}_B(6,J,\rho) \nonumber\\
&=&\frac{1}{8} \left(2 \mathcal{W}_{0,2}+\mathcal{W}_{3,0}-8\tilde{\alpha}_0(6)-700\tilde{\alpha}_2(6)-6552\tilde{\alpha}_4(6) \right) \nonumber\\ &+&\frac{1}{8} \left(4 \mathcal{W}_{0,2}-3\mathcal{W}_{3,0}-6\tilde{\alpha}_0(6)+210\tilde{\alpha}_2(6)-3654\tilde{\alpha}_4(6) \right)  \rho \nonumber\\&+& \frac{1}{8} \left(2 \mathcal{W}_{0,2}+3 \mathcal{W}_{3,0}-12\tilde{\alpha}_0(6)+120\tilde{\alpha}_2(6)-1548\tilde{\alpha}_4(6) \right)  \rho^2 \nonumber \\ &+&\frac{1}{8} \left(-\mathcal{W}_{3,0}+2\tilde{\alpha}_0(6)+10\tilde{\alpha}_2(6)+18\tilde{\alpha}_4(6) \right)  \rho^3 \,, 
\end{eqnarray}
gives 
\begin{eqnarray}
\mathcal{S}(6)&:& 3 \tilde{\alpha} _0(6)-75 \tilde{\alpha} _2(6)+747 \tilde{\alpha} _4(6)\leq \mathcal{W}_{0,2} \le 3\tilde{\alpha} _0(6)-45 \tilde{\alpha} _2(6)+927
   \tilde{\alpha} _4(6), \nonumber\\ && \left(2 \tilde{\alpha} _0(6)+10 \tilde{\alpha}_2(6)+18 \tilde{\alpha}
   _4(6)\right) \le \mathcal{W}_{3,0} \le \left(-2 \mathcal{W}_{0,2}+8 \tilde{\alpha} _0(6)+700\tilde{\alpha} _2(6)+6552 \tilde{\alpha}
   _4(6) \right) \nonumber\\  TR(6)&:& 3 \tilde{\alpha} _0(6)-75 \tilde{\alpha} _2(6)+747 \tilde{\alpha} _4(6)\leq \mathcal{W}_{0,2} \le 3\tilde{\alpha} _0(6)-55 \tilde{\alpha} _2(6)+867
   \tilde{\alpha} _4(6), \nonumber\\ && \mathcal{W}_{3,0} \ge \frac{1}{3}\left(2 \mathcal{W}_{0,2}+120 \tilde{\alpha} _2(6)-1800\tilde{\alpha}
   _4(6) \right) \nonumber\\  \mathcal{D}(6)&:&|\tilde{R}_T(6,\rho)|\le \frac{1}{4} \left(1+\sqrt{5}\right)^2\approx 2.618
\end{eqnarray}

All of which are obeyed by the string case. Thus, we can get {\bf non-projective bounds} \cite{yutin3,yutin4} such as the ones described above for any $\mathcal{W}_{p,q}$ with $2p+3q\ge 2$. We shall briefly address the two cases namely $\mathcal{W}_{1,0},\mathcal{W}_{0,1}$ where the bounds do not follow from typically realness since the $\tilde{R}_T(\rho)=\rho$ in these cases. From \eqref{fbex} it follows that:
\begin{eqnarray}
  \mathcal{W}_{1,0} &=& 2\sum_{J=0}^{\infty} (2J+1) \tilde{\alpha}_J(2) \ge 2 \tilde{\alpha}_0(2)+ 10 \tilde{\alpha}_2(2) \ge 0  \nonumber\\
  \mathcal{W}_{0,1} &=& \sum_{J=0}^{\infty} (2J+1) \tilde{\alpha}_J(3) (2 J^2+2J-3) \ge -3 \tilde{\alpha}_0(3)
\end{eqnarray}
one could the upper bounds by using the facts $\tilde{\alpha}_J(n)\ge \tilde{\alpha}_J(n+1)$ and $\tilde{\alpha}_0(4)>  \lambda \tilde{\alpha}_J(4)$ where\footnote{This follows from \eqref{lsd} for $J=0$ and \eqref{alphatn4bound1} for $J\ge2$.} $\lambda=10$. This gives $2 \tilde{\alpha}_0(2)+ 10 \tilde{\alpha}_2(2) \le \mathcal{W}_{1,0} \le \tilde{\alpha}_0(1) \sum_{J=0}^{\infty} (2J+1)(0.1)^{-J/2} \le \frac{25.67}{M^2} $ and $-3 \tilde{\alpha}_0(3) \le \mathcal{W}_{0,1} \le \tilde{\alpha}_0(2) \sum_{J=0}^{\infty} (2J+1)(2J^2+2J-3)(0.1)^{-J/2} \le \frac{24.71}{M^4}   $.
Thus we have 
$$ 0 \le \mathcal{W}_{1,0}\le \frac{25.67}{M^4},-\frac{4}{M^6} \le \mathcal{W}_{0,1}\le \frac{24.71}{M^6} $$
\noindent One can also readily get projective bounds by comparing partial wave moments at different orders since $\tilde{\alpha}_J(n) > \tilde{\alpha}_J(n+1)$ identically to the way we obtained upper bounds for $\mathcal{W}_{1,0},\mathcal{W}_{1,1}$ above. The interested reader may look at eq(\ref{w20w11}) in app(\ref{sec:dposexamples}) for an example.

We could also estimate the error when we  truncate the partial wave expansion to $J<J_{max}$ and we bound this quantity and this is most easily done at the level of \eqref{Rtn1}. We demonstrate this for $n=4,5$ cases below:
\begin{eqnarray}
{\bf n=4:}&& \Bigg|\frac{R_T(4,\rho)+\frac{1}{4} (2 \tilde{\alpha}_0(4)+190 \tilde{\alpha}_2(4)-\mathcal{W}_{2,0})}{\frac{1}{2}\left(2\tilde{\alpha}_0(4)-20\tilde{\alpha}_2(4)-\mathcal{W}_{2,0} \right)}\Bigg|\le 1.707 \nonumber\\
&\implies& \bigg|\sum_{J=J_T}^{\infty} (2J+1)\tilde{\alpha}_J \mathcal{F}_B(4,J,\rho)\bigg|\le f_1(\tilde{\alpha}_0(4),\tilde{\alpha}_2(4),\mathcal{W}_{2,0}) \\
{\bf n=5:}&& \Bigg|\frac{R_T(5,\rho)+\frac{1}{4} (5 \tilde{\alpha}_0(5)+265 \tilde{\alpha}_2(5)+\mathcal{W}_{1,1})}{\frac{1}{4}\left(5\tilde{\alpha}_0(5)-35\tilde{\alpha}_2(5)+\mathcal{W}_{1,1} \right)}\Bigg|\le 1.707 \nonumber\\
&\implies& \bigg|\sum_{J=J_T}^{\infty} (2J+1)\tilde{\alpha}_J \mathcal{F}_B(n,J,\rho)\bigg|\le f_2(\tilde{\alpha}_0(5),\tilde{\alpha}_2(5),\mathcal{W}_{1,1})
\end{eqnarray}

where,
\begin{eqnarray}
    f_1(\tilde{\alpha}_0(4),\tilde{\alpha}_2(4),\mathcal{W}_{2,0})&=&Max \left(1.207\tilde{\alpha}_0(4) - 64.57 \tilde{\alpha}_2(4) + 1.103 \mathcal{W}_{20}, \right. \nonumber\\ 
    && \left. 2.207 \tilde{\alpha}_0(4) + 30.43 \tilde{\alpha}_2(4) +0.6035 \mathcal{W}_{20} \right)\,,\\f_2(\tilde{\alpha}_0(5),\tilde{\alpha}_2(5),\mathcal{W}_{1,1})&=& Max \left(0.883\tilde{\alpha}_0(5) - 81.1863 \tilde{\alpha}_2(4) + 0.176 \mathcal{W}_{20}, \right. \nonumber\\ 
    && \left. 3.384 \tilde{\alpha}_0(4) + 51.314 \tilde{\alpha}_2(4) +0.676 \mathcal{W}_{20} \right)\,.
\end{eqnarray}

For the string case the above values are $f_1=3.46351,f_2=0.0457$.
%%%%%%%%%%%%%%%%%%%%%%%%%%%%%%%%%%%%%%%%%%%%%%%%%%%%%%%%%%%%%%%%%%%%%%%%
\subsection{Massless poles}
We now consider amplitudes with massless poles. We could address these cases by considering the class of meromorphic typically real polynomials of degree $N$ called $TM^N$ which are the polynomial analogues of the Goodman class \cite{goodman} $TM^*$ considered in \cite{rs1} and are of the form $f(z)=\frac{1}{z}+\sum_{i=0}^N a_i z^i$. However this is beyond the scope of the current paper and for now we take the simplified approach of looking at  the coefficient bounds one obtains by assuming the rest of the low energy expansion is typically real.
\subsubsection{The scalar pole}
We consider an amplitude $M(\tilde{\rho})$ with a massless scalar \footnote{This was a case that could not be directly addressed in \cite{rs1} as $\frac{x}{y}=\frac{1}{a}$ which is  constant and thus independent of the Auberson-Khuri $z$ variable.} exchange $-\left(\frac{1}{s}+\frac{1}{t}+\frac{1}{u}\right)=-\frac{x}{y}$:
\begin{eqnarray}
M(\tilde{\rho}) &=&   -\lambda \frac{x}{y}+ \mathcal{W}_0~ x+ \mathcal{W}_1 ~y\cdots \nonumber \\
&=&  \frac{\lambda}{\omega^2} \frac{(2-\tilde{\rho})}{\tilde{\rho}}+ \mathcal{W}_0~ \omega^4 \frac{(2-\tilde{\rho})}{2}-\mathcal{W}_1 ~ \omega^6~ \frac{\tilde{\rho}}{2}+\cdots
\end{eqnarray}
We re-scale the amplitude and introduce a change of variables $\tilde{\rho}=z~ \alpha$ with $|z|<1$, $\alpha>0$. The parameter $\alpha$ is the largest disk around the origin for which the amplitude is a typically real meromorphic function in $\tilde{\rho}$. We map this on to the unit disk in $z$ since most of the results in \cite{goodman,rs1} are valid for this domain. 
\begin{eqnarray}
-\frac{\alpha ~\omega^2 }{2~ \lambda}M(z) &=& -\frac{1}{z}+ \alpha \left(- \frac{\hat{\mathcal{W}_0}~ \omega^6}{2}+1\right)+ \frac{\alpha^2 \omega^6 }{4}~(\hat{\mathcal{W}_1} ~\omega^2 +\hat{\mathcal{W}_0})  z+ \cdots
\end{eqnarray}
where, $\hat{\mathcal{W}_i}=\frac{\mathcal{W}_i}{\lambda}$. Having done this we can apply the inequalities of Goodman for the function in $TM^*$ to get:
\begin{eqnarray}
    \alpha \left( \frac{-\hat{\mathcal{W}_0} \omega^6}{2}+1\right) &\ge& -1 \,, \nonumber \\
    \implies~~~~~~~~~~~~~~ \hat{\mathcal{W}_0} &\le& \frac{(1+\alpha)}{\alpha~ \omega^6}
\end{eqnarray}
Furthermore as explained in \cite{goodman,rs1},  $f(z) \in TM^*$ iff $-\frac{1}{f(z)} \in TM$. By assuming the smallest non-zero pole is at $|z|=p>0$ (this corresponds to the lightest massive particle in the theory) and applying the Goodman bounds we get :
\begin{eqnarray}
    \frac{2~ \lambda}{\alpha ~\omega^2~ M(z)} &=& z+ \alpha^2\left(1-\hat{\mathcal{W}_0}\omega^6 \right) z^2+\cdots  \nonumber\\
    \implies~~~~~~~~~~~~~~~~ -\left(p+\frac{1}{p} \right)\le \alpha^2\left(1-\hat{\mathcal{W}_0}\omega^6 \right) &\le& \left(p+\frac{1}{p} \right) \nonumber\\
  -\frac{1}{\omega^6}\left(\frac{1}{\alpha^2}\left(p+\frac{1}{p}\right)-1\right)  \le\hat{\mathcal{W}_0} &\le& \frac{1}{\omega^6}\left(\frac{1}{\alpha^2}\left(p+\frac{1}{p}\right)+1\right)
\end{eqnarray}
 Thus we have 
 \begin{equation}
\boxed{ -\frac{p^2 -p \alpha^2+1}{p~ \alpha^2~\omega^6}\le \hat{\mathcal{W}_0} \le Min\bigg{\{} \frac{(1+\alpha)}{\alpha~ \omega^6},\frac{p^2 +p \alpha^2+1}{p~ \alpha^2~\omega^6}\bigg{\}} }\,.
\end{equation}
For the terms till $\omega^6$ we have $\alpha=2$ and $p=1$ which gives $\frac{1}{2 \omega^6}\le \hat{\mathcal{W}_0}\le \frac{3}{2 \omega^6}$. 

For $\alpha=1$ and $p=1$ we have $-\frac{1}{\omega^6}\le \hat{\mathcal{W}_0}\le \frac{2}{\omega^6}$. 
Since $0<p\le 1$ and $\alpha>0$ this suggests that $\hat{\mathcal{W}_0}\ge 0$ for $\alpha \ge \sqrt{p+1/p}$. For the case where there are no non-trivial poles i.e $p=1$ this suggests that we will have $\alpha\ge \sqrt{2}\approx 1.414$ since $\hat{\mathcal{W}_0}\ge 0$ is implied by unitarity.

\subsubsection{The graviton pole}
We consider the $\frac{1}{s t u}$ SUGRA pole in 10D along with the $R^4$ term. 
\begin{eqnarray}
M(\tilde{\rho}) &=&   - 8 \pi G_N \frac{x^2}{y}+ g_0 x^2+ \cdots \nonumber \\
&=&  4 \pi G_N \omega^2 \frac{(2-\tilde{\rho})^2}{\tilde{\rho}}+ g_0 \omega^8 \frac{(2-\tilde{\rho})^2}{4}
\end{eqnarray}

We analyze the coefficient bounds one obtains for $g_0$ by assuming the rest of the low energy expansion is typically real. To do this as before we first re-scale the amplitude and change variables to $\tilde{\rho}= z~ \alpha$ as done in the scalar case we get
\begin{eqnarray}
\frac{-M(z)}{ 16 \pi G_N \omega^2} &=& -\frac{1}{z}+ \left(\alpha -\frac{1}{2} \alpha ~ \hat{g_0} ~\omega^6\right)+\left( \frac{1}{2} \alpha ^2 ~\hat{g_0} ~\omega^6 -\frac{\alpha ^2}{4}\right) z -\frac{1}{8} \alpha ^3 ~\hat{g_0}~ \omega^6 z^3+ \cdots
\end{eqnarray}

 where, $\hat{g_0}=\frac{g_0}{8\pi G_N}$. The change of variables to $z$ is needed since the Goodman bounds hold for a disc of unit radius. As noted earlier, $M(\tilde\rho)$ to $O(\omega^8)$ is TR in a disc with $\alpha=2$. Strictly speaking, this is a weak-coupling result since we have ignored terms proportional to $\log(s)$ in the expansion. However, now let us assume that the full amplitude, even at strong-coupling, $M(\tilde\rho)$ is TR inside $|z|<1$ and examine the consequences. $M(\tilde\rho)$ above is an element of Goodman class $TM^{*}$ as discussed in  \cite{rs1} and hence satisfies :
\begin{eqnarray}
   && \left( \frac{1}{2} \alpha ^2 ~\hat{g_0} ~\omega^6 -\frac{\alpha ^2}{4}\right) \ge -1 \nonumber \\
   &\implies& \hat{g_0} \ge \frac{(\alpha^2-4)}{2 ~\omega^6 ~\alpha^2}
\end{eqnarray}

Furthermore as discussed in \cite{rs1} we can convert this to an element of the Goodman class $TM$ and by assuming that there are no poles inside the disk $|z|<p$ we get the following coefficient  bound:
\begin{equation}
     \hat{g_0} \le \frac{2}{\omega^6 \alpha} \left(p+\frac{1}{p}+\alpha \right)
\end{equation}
 Thus we have 
 \begin{equation}
\boxed{\frac{(\alpha^2-4)}{2 ~\omega^6 ~\alpha^2}\le \hat{g_0} \le \frac{2}{\omega^6 \alpha} \left(p+\frac{1}{p}+\alpha \right)}\,.
\end{equation}

This is a two-sided bound on $\hat g_0$ and is does not explicitly depend on the spacetime dimensions. Notice that the upper bound depends on both $p,\alpha$ while the lower bound only depends on $\alpha$. So far in the literature \cite{sch2, rs1}, the lower bound in weakly coupled theories is 0. Unitarity tells us that $\hat g_0\geq 0$ \cite{pedro} which tells us that $\alpha\geq 2$.

The derivation we have presented above only assumes TR-ness of the full amplitude up to a certain value $\omega^2$. 
The values $p=1$ and $\alpha=2$ would give us the bound obtained in \cite{rs1} namely $0\le \hat{g_0}\le 4$. The values $p=1$ and $\alpha=2.32$  gives us the bound $0.13 \lesssim \hat{g_0}\lesssim 3.7$, where the lower bound coincides with the strong string coupling result obtained in \cite{pedro}. This suggests that the the size of the disc $\alpha$ where TR-ness holds is sensitive to the string coupling and $\alpha(g_s=0)=2$ while\footnote{The location of the nearest pole to the origin $p$ in the above formula presumably goes to zero as the string coupling increases---this removes the upper bound at strong string coupling \cite{sch2}. } $\alpha(g_s=\infty)\approx 2.32$.
%%%%%%%%%%%%%%%%%%%%%%%%%%%%%%%%%%%%%%%%%%%%%%%%%%%%%%%%%%%%%%%

\section{Discussion}
We will now conclude with a brief discussion of promising future directions of research. %{\bf SG, PR: Have a go at this. Itemize 4-5 interesting directions that you can think of.}

\begin{itemize}
    \item {\bf Celestial OPE:} It is natural to ask what one can learn about celestial conformal field theories (CCFT's) from the 4d S-matrix. In the context of the results obtained in this paper, one would wonder if there is a relation between the partial wave coefficients of the momentum space amplitude and OPE coefficients in the $2$-$d$ conformal block decomposition of the corresponding celestial amplitude. It turns out that this is indeed the case and will be the topic of our upcoming work \cite{toappear}.   
    
    \item {\bf External gravitons/spinning particles:} One would also like to extend the analysis of this paper to cases with external spinning particles. In particular, for the case with external gravitons is of particular interest in the context of CCFT's. It will be also be interesting to see the implications of null constraints for LSD in this case and compare with existing literature\cite{sasha}.This is ongoing work using techniques developed in \cite{rs2} and we hope report progress on this front soon. 
    
    \item {\bf External massive:} It is also of great interest to see if the techniques of this paper can be extended to external massive particles both from the CCFT perspective and also from the perspective of the 4d S-matrix since it would be fascinating to see if the analogues of the partial wave moment bounds we obtained in this work could help address the presence/absence of LSD in this context (see \cite{rs2} for a related discussion). We leave this for future work.
    
    \item {\bf Analytic structure for general $\beta$:} It is also interesting to consider the analytic structure  of the celestial amplitude for general $\beta$ since we have focused mainly on the low energy regime i.e., $\beta=-2 n$ in this work. It would be interesting to see if the high energy regime $\beta= 2 n$ can be addressed similarly \cite{Arkani-Hamed:2020gyp}. We can ask if the positivity properties we have considered persist for general values of $\beta$ and if these lead to any non-trivial consequences for the CCFT. This merits further study. %More generally since the celestial amplitude is just a different way of writing the 4-d S-matrix and thus contains the same information as the momentum space amplitude, we would like to better understand what information about the amplitude is encoded in generic values of $\beta \neq 2 \mathbb{Z}$.%)
    
    \item {\bf Positivity:} Finally it is also a fascinating mathematical question to better explore connection between the notions of positivity introduced in this work and those studied in the maths literature such as Toeplitz positivity. The Feynman blocks introduced in this work were a family of typically-real polynomials and it would be interesting to better understand the connection between the Suffridge polynomials and the Feynman blocks. 
\end{itemize}

\section*{Acknowledgements}
We thank Faizan Bhat, Parthiv Haldar and Ahmadullah Zahed for discussions. We thank Shamik Banerjee for valuable comments on the draft. S.G. is supported by a Raman postdoctoral position of IISc while P.R. is supported by an IOE endowed postdoctoral position at IISc.
A.S. acknowledges support from MHRD, Govt. of India, through a SPARC grant P315 and from DST through the SERB core grant CRG/2021/000873. 

\appendix

%%%%%%%%%%%%%%%%%%%%%%%%%%%%%%%%%%%%%%%%%%%%%%%%%%%%%%%%%%%%%%%
\section{Some proofs}
\label{sec:positivityproofs}

In this section we prove the $\rho$-positvity $\mathcal{P}_\rho$ of the amplitudes listed in section (\ref{sec:newpositivity}). We begin with the type II amplitude which we write in terms of an exponential involving zeta functions.
 \begin{eqnarray}
-\mathcal{M}_{II}&=& \frac{-1}{s t u} \frac{\Gamma(1-s)~\Gamma(1-t)~\Gamma(1-u)}{\Gamma(1+s)~\Gamma(1+t)~\Gamma(1+u)} \nonumber\\
&=& \frac{2}{\omega^6 (1+\rho)} e^{- 8 \gamma}~ \exp\left( \sum_{k=2}^{\infty} \frac{\zeta(k) \omega^{2 k}}{k} \left(1-(-1)^k\right) \left(s^k+t^k+u^k\right)\right)\nonumber\\
&=& \frac{2}{\omega^6 (1+\rho)} e^{- 8 \gamma}~ \exp\left( \sum_{{k=3}\atop{k~ odd}}^{\infty} \frac{2 \zeta(k)}{k} f(k,\rho,\omega)\right)\,.
 \end{eqnarray}
 
where  $f(k,\rho,\omega)= \omega^k(1+(-z)^k+(z-1)^k)$, and $z= \frac{1}{2}\left(1-\sqrt{-1-2\rho}\right)$. The $f(k,\rho,\omega)$'s are manifestly positive for {\it any} $k$ as they satisfy the following 3-term recursion relation with positive coefficients as can be readily checked.
\begin{eqnarray} \label{identityf}
f(k,\rho,\omega)&=&\frac{1}{2} \omega^6~(1+\rho) f(k-3,\rho,\omega)+ \frac{1}{2} \omega^4~(1-\rho) f(k-2,\rho,\omega)~ for~ k\ge 3
\end{eqnarray}

with $f(0,\rho,\omega)=3,f(1,\rho,\omega)=0$ and $f(2,\rho,\omega)=16+ \omega^4(1-\rho)$. Positivity is now obvious to see since the functions $1\pm\rho$ are both positive in $\rho\in(-1,1)$ and the we can recursively construct any $f(k,\rho,\omega)$ for $k\ge3$ using the above relation which makes positivity manifest.

 Now let us consider the closed bosonic string amplitude. This can also be written in an exponential form as follows:
\begin{eqnarray}
-\mathcal{M}_{CB}&=& \frac{\Gamma\left(1-\left(\frac{2}{3}+s_1\right)\right)~\Gamma\left(1-\left(\frac{2}{3}+s_2\right)\right)~\Gamma\left(1-\left(\frac{2}{3}+s_3\right)\right)}{\Gamma\left(1-\left(\frac{1}{3}-s_1\right)\right)~\Gamma\left(1-\left(\frac{1}{3}-s_2\right)\right)~\Gamma\left(1-\left(\frac{1}{3}-s_3\right)\right)} \nonumber\\
&=& \exp\left( \gamma+ \sum_{k=2}^{\infty} \frac{\zeta(k)}{k} \sum_{i=1}^3 \left( \left(s_i +\frac{2}{3}\right)^i-\left(-s_i +\frac{1}{3}\right)^i\right) \right) \nonumber\\
&=& e^\gamma \exp\left( \sum_{k=2}^{\infty} \frac{\zeta(k)}{k} \sum_{n=0}^{k-1} \lambda(n,k) f(k,\rho,\omega) \right) 
\end{eqnarray}

where $\lambda(n,k)=\frac{2^{-n} 3^{n-k} \left(2^n \left(n (1-k)_{n-1}-(1-k)_n\right)+2^k
(k-n+1)_n\right)}{n!}$ and $f(k,\omega,\rho)=s_1^k+s_2^k+s_3^k$ is the same function defined in the type II case. Positivity in $\rho \in (-1,1)$ readily follows as $f(k,\omega,\rho)$ were argued above to be positive and $\lambda(n,k)$ are manifestly positive.   

One can show that this positivity is also enjoyed by tree-level and 1-loop box diagram for 2-2 scattering in $\phi^2 \psi$ theory, with  $\phi$ being a light scalar and $\psi$ a heavy one. These amplitudes are respectively given by,
\begin{eqnarray}
-\mathcal{M}_{\phi^3}&=& g\left(\frac{1}{m^2-s}+\frac{1}{m^2-t} +\frac{1}{m^2-u}  \right) \nonumber\\
&=& g \sum_{k=0}^{\infty} \frac{f(k,\rho,\omega)}{m^{2k}} \\
-\mathcal{M}_{\phi^2 \psi}&=& \frac{\pi^2}{6 m^4}\left( F_3 \left({{1,1,1,1}\atop{\frac{5}{2}}}; \frac{s}{4 m^2},\frac{t}{4 m^2} \right)+(s \rightarrow t,t \rightarrow u) +(s \rightarrow u,t \rightarrow s)\right) \nonumber \\
&=&\frac{\pi^2}{6 m^4}\left( \sum_{p,q=0}^{\infty} \frac{p!~q!}{\left(\frac{5}{2}\right)_{p+q} (4~m^2)^{p+q}}\frac{\left(f(p,\rho,\omega)f(q,\rho,\omega)-f(p+q,\rho,\omega)\right)}{2}\right) \nonumber\\
&=&\frac{\pi^2}{6 m^4}\left( \sum_{n=0}^{\infty}\sum_{p=0}^n \frac{p!~(n-p)!}{\left(\frac{5}{2}\right)_{n} (4~m^2)^{n}}\frac{\left(f(p,\rho,\omega)f(n-p,\rho,\omega)-f(n,\rho,\omega)\right)}{2}\right) 
\end{eqnarray}

%The positivity of the former follows from the positivity of $f(k,\rho,\omega)$. 
The positivity follows as $g(n)=\sum_{p=0}^n p! ~(n-p)! \left(f(p,\rho,\omega)f(n-p,\rho,\omega)-f(n,\rho,\omega)\right) \ge 0$ for $n\le 2$ as $g(0)=6,g(1)=0,g(2)=7 f(2,\rho,\omega)$ can be easily seen and proved inductively for $n \ge 3$ by using \eqref{identityf}. 
%Since,  
%\small{\begin{equation}
%g(n)= \frac{ \omega^{6}}{2}~(1+\rho) g(n-3)+ \frac{\omega^4}{2} ~(1-\rho) g(n-2)+ ((2n-1)(n-1)!+1)f(n,\rho,\omega)\\
%\end{equation}
%\noindent can be used to express any $g(p,q)$ as a positive sum of lower $g$'s. 

{\it Thus, we see that positivity of the low energy coefficients for $\rho \in (-1,1)$ follows purely from the positivity of $f(k,\rho,\omega)$. }

%%%%%%%%%%%%%%%%%%%%%%%%%%%%%%%%%%%%%%%%%%%%%%%%%%%%%%%%%%%%%%%%%%%%
\section{ $4$-point celestial amplitude of massless particles} 
\label{sec:csampreview}

In this appendix we review the construction of $4$-point celestial amplitudes for massless particles in $d$=4 spacetime dimensions. 

The null four-momenta of the external particles in the S-matrix can be parametrized as 
\begin{equation}
\label{nullmom}
\begin{split}
p^{\mu}_{k} = \epsilon_{k} \omega_{k} (1+ z_{k} \bar{z}_{k}, z_{k}+ \bar{z}_{k}, -i (z_{k}-\bar{z}_{k}), 1-z_{k}\bar{z}_{k})
\end{split}
\end{equation}

where $\epsilon_{k} =\pm 1$ for an outgoing (incoming) particle. $\omega_{k}$ is the energy of the $k$-th particle. $(z_{k},\bar{z}_{k})$ specify the direction of the null momenta $p^{\mu}_{k}$ and thereby can be identified with stereographic coordinates on the $2$-$d$ celestial sphere at null infinity.  

\vskip 4pt
Now the $4$-point scattering amplitude in momentum space can be expressed as 
\begin{equation}
\label{4ptamp}
\begin{split}
\mathbf{A}(p_{1},p_{2},p_{3},p_{4}) = \prod_{i< j}^{4} \left(\frac{z_{ij}}{\bar{z}_{ij}} \right)^{\frac{1}{2}\left(\frac{J}{3}- J_{i} - J_{j} \right)} \ \mathbf{M}(s,t); \quad J = \sum_{i=1}^{4} J_{i}
\end{split}
\end{equation}

where $J_{i}$ denotes the helicity of the $i$-th particle.  The $(z_{i}, \bar{z}_{i})$ dependent prefactor in \eqref{4ptamp} accounts for the Lorentz transformation properties of the amplitude. $\mathbf{M}(s,t)$ is a Lorentz-invariant function of the Mandelstam invariants $s=-(p_{1}+p_{2})^{2}$ and $t= -(p_{1}+p_{3})^{2}$. 

\vskip 4pt

The celestial amplitude corresponding to \eqref{4ptamp} is then defined as
\begin{equation}
\label{4ptcsamp1}
\begin{split}
\mathcal{M}(\Delta_{i}, J_{i}, z_{i},\bar{z}_{i}) & =  \int_{0}^{\infty} \prod_{i=1}^{4} d\omega_{i} \ \omega_{i}^{\Delta_{i}-1} \ \mathbf{A}(p_{1},p_{2},p_{3},p_{4}) \ \delta^{(4)}\left(\sum_{i=1}^{4}p^{\mu}_{i}\right)
\end{split}
\end{equation}

Now note that under $4$-$d$ Lorentz transformations, the $(\omega_{i},z_{i},\bar{z}_{i})$ variables transform as follows
\begin{equation}
\label{zlt}
\begin{split}
\omega_{i} \rightarrow |cz_{i}+d|^{2} \omega_{i}; \quad (z_{i},\bar{z}_{i}) & \rightarrow \left(\frac{az_{i}+b}{cz_{i}+d},\frac{\bar{a}\bar{z}_{i}+\bar{b}}{\bar{c}\bar{z}_{i}+\bar{d}}  \right); \quad ad-bc= \bar{a}\bar{d}- \bar{b}\bar{c} =1
\end{split}
\end{equation}

Using this it can be shown that under the action of the $4$-$d$ Lorentz group which is isomorphic to $SL(2,\mathbb{C})$,  the celestial amplitude \eqref{4ptcsamp1} transforms as
\begin{equation}
\label{4pointmellin}
\begin{split}
\mathcal{M}(\Delta_{i}, J_{i}, z_{i},\bar{z}_{i}) & \rightarrow \prod_{i} (c{z}_{i}+d)^{-\Delta_{i}-J_{i}} (\bar{c}\bar{z}_{i}+\bar{d})^{-\Delta_{i}+J_{i}} \mathcal{M}\left(\Delta_{i}, J_{i}, \frac{az_{i}+b}{cz_{i}+d},\frac{\bar{a}\bar{z}_{i}+\bar{b}}{\bar{c}\bar{z}_{i}+\bar{d}}\right)
\end{split}
\end{equation}

Thus $\mathcal{M}(\Delta_{i}, J_{i}, z_{i},\bar{z}_{i})$ transforms in the same fashion under $SL(2,\mathbb{C})$ as a $4$-point correlation function of quasi-primary operators  with scaling dimensions $\Delta_{i}$ and spins $J_{i}$ in $2$-$d$ CFT, referred to as the Celestial CFT (CCFT).  

\vskip 4pt
Now \eqref{4ptcsamp1} can be further simplified as follows. Using the momentum-conservation delta function we can perform the integrals over any $3$ of the $\omega_{i}$ variables. For this it is convenient to write the delta function as
\begin{equation}
\label{deltarep}
\begin{split}
\delta^{(4)}\left(\sum_{i=1}^{4}p^{\mu}_{i}\right) = \frac{1}{4\omega_{4}^{2}} \ \frac{\delta(z-\bar{z})}{|z_{12}\bar{z}_{12}z_{34}\bar{z}_{34}|} \prod_{i=1}^{3} \delta(\omega_{i}-\epsilon_{i}\epsilon_{4}\omega_{4} \sigma_{i}) \prod_{j=1}^{3}\Theta(\epsilon_{j}\epsilon_{4} \sigma_{j})
\end{split}
\end{equation}

where $(z,\bar{z})$ are the cross-ratios
\begin{equation}
\label{zzbardef}
\begin{split}
z =\frac{z_{13}z_{24}}{z_{12}z_{34}}; \quad \bar{z} = \frac{\bar{z}_{13}\bar{z}_{24}}{\bar{z}_{12}\bar{z}_{34}}
\end{split}
\end{equation}

and
\begin{equation}
\label{sigma}
\begin{split}
& \sigma_{1} = - z \ \frac{z_{34}\bar{z}_{34}}{z_{13}\bar{z}_{13}}; \quad \sigma_{2} =   (z-1) \ \frac{z_{34}\bar{z}_{34}}{z_{23}\bar{z}_{23}}; \quad \sigma_{3} = -  \frac{z}{z-1} \ \frac{z_{14}\bar{z}_{14}}{z_{13}\bar{z}_{13}}
\end{split}
\end{equation}

The theta functions in \eqref{deltarep} ensure that the delta functions have support only for $\omega_{i} >0$. Consequently the cross ratio $z$ lies in the following ranges depending on the scattering channel\footnote{A scattering channel here refers to a particular configuration of in/out states.} under consideration. 
\begin{equation}
\label{zrange}
\begin{split}
& 12 \leftrightarrow 34 \ \text{channel}: \epsilon_{1} = \epsilon_{2} = - \epsilon_{3} = -\epsilon_{4} = -1; \quad z \in (0,1) \\
& 13 \leftrightarrow 24 \ \text{channel}: \epsilon_{1} = \epsilon_{3} = - \epsilon_{2} = -\epsilon_{4} = -1; \quad z \in (1,\infty) \\
& 14 \leftrightarrow 23 \ \text{channel}: \epsilon_{1} = \epsilon_{4} = - \epsilon_{2} = -\epsilon_{3} = -1; \quad z \in (-\infty, 0)
\end{split}
\end{equation}

In terms of the $(\omega_{i},z_{i},\bar{z}_{i})$ variables, the Mandelstam invariants $s,t$ become
\begin{equation}
\label{stvals}
\begin{split}
& s = -(p_{1}+p_{2})^{2} = - 4 \omega_{4}^{2} \ \frac{z}{z-1} \ \frac{z_{14}\bar{z}_{14}z_{34}\bar{z}_{34}}{z_{13}\bar{z}_{13}} , \quad  t= - (p_{1}+p_{3})^{2} = 4 \omega_{4}^{2} \ \frac{z^{2}}{z-1} \ \frac{z_{14}\bar{z}_{14}z_{34}\bar{z}_{34}}{z_{13}\bar{z}_{13}}
\end{split}
\end{equation}

where we have evaluated \eqref{stvals} on the support of the $\delta(z-\bar{z})$ factor. Then \eqref{4ptcsamp1} takes the form
\begin{equation}
\label{4ptmellin2}
\begin{split}
\mathcal{M}(h_{i},\bar{h}_{i}, z_{i},\bar{z}_{i}) & =    \frac{\delta(z-\bar{z})}{4 |z_{12}\bar{z}_{12}z_{34}\bar{z}_{34}|} \prod_{i=1}^{3}\left(\epsilon_{i}\epsilon_{4} \sigma_{i}\right)^{\Delta_{i}-1}  \prod_{j=1}^{3}\Theta(\epsilon_{j}\epsilon_{4} \sigma_{j})  \prod_{k< l}^{4} \left(\frac{z_{kl}}{\bar{z}_{kl}} \right)^{\frac{1}{2}\left(\frac{J}{3}- J_{k} - J_{l} \right)}  \\ 
&   \times \int_{0}^{\infty}  d\omega_{4} \ \omega_{4}^{\beta-1} \ \mathbf{M}(s,t) 
\end{split}
\end{equation}

Next we perform the change of variables $\omega = \sqrt{\epsilon_{s} s}$ where $\epsilon_{s}=\epsilon_{1}\epsilon_{2}$. Note that $\epsilon_{s}=1$ in $12\leftrightarrow 34$ channel and $\epsilon_{s}=- 1$ in $13\leftrightarrow 24$  and $14\leftrightarrow 23$ channels respectively. Then \eqref{4ptmellin2} can be shown to take the following form
\begin{equation}
\label{4ptmellin3}
\begin{split}
\mathcal{M}(h_{i},\bar{h}_{i}, z_{i},\bar{z}_{i})  = K_{h_{i},\bar{h}_{i}}(z_{i},\bar{z}_{i}) \  X(\beta, z) \ \int_{0}^{\infty}  d\omega \ \omega^{\beta-1} \ \mathbf{M}(\epsilon_{s}\omega^{2}, - \epsilon_{s}  \omega^{2}z) 
\end{split}
\end{equation}

where
\begin{equation}
\label{KXdef}
\begin{split}
& K_{h_{i},\bar{h}_{i}}(z_{i},\bar{z}_{i}) = \prod_{i<j}^{4} z_{ij}^{\frac{h}{3}-h_{i}-h_{j}} \bar{z}_{ij}^{\frac{\bar{h}}{3}-\bar{h}_{i}-\bar{h}_{j}}, \quad h=\sum_{i=1}^{4}h_{i}, \quad \bar{h} = \sum_{i=1}^{4}\bar{h}_{i}  \\
& X(\beta, z) = 2^{-\beta-2}  \ |z(1-z)|^{\frac{(\beta+4)}{6}} \ \delta(z-\bar{z}) \prod_{j=1}^{3}\Theta(\epsilon_{j}\epsilon_{4} \sigma_{j}), \quad \beta = \sum_{i=1}^{4}(\Delta_{i}-1)  %\   \epsilon_{s}^{\beta/2} \prod_{i=1}^{3}\left(\epsilon_{i}\epsilon_{4} \right)^{\Delta_{i}-1} \ (-1)^{\Delta_{1}+\Delta_{2}}
\end{split}
\end{equation}

%%%%%%%%%%%%%%%%%%%%%%%%%%%%%%%%%%%%%%%%%%%%%%%%%%%%%%%%%%%

\section{Some explicit formulas}
\label{sec:formulas}
In this appendix we give closed form expressions for some quantities used in the main text in sections (\ref{sec4}) and (\ref{sec5}).
We can simplify eq (\ref{Rjdef}) for $\mathcal{R}_{J}(\beta, x_{i}(z))$ to obtain the following: 
\begin{equation}
  \mathcal{R}_{J}(\beta, x_{i})= \sum_{k=0}^{J/2-1}  \mathcal{S}(k,\beta,x_i) \mathcal{T}(J,J/2-k-1)\, ,
\end{equation}
where,
{\small{\begin{equation}
\begin{split}
& \mathcal{S}(k,\beta,x_i)= -e^{\frac{i \pi  \beta }{2}} (x_i+1)^{-k-1} \left((-1)^k (x_i+1)^{k+1}
   \binom{\frac{\beta }{2}}{k+1} \, _2F_1\left(1,k-\frac{\beta
   }{2}+1;k+2;x_i+1\right)+(-x_i)^{\beta /2}\right) \,,\\
  &  \mathcal{T}(J,n)= \frac{3^n 2^{2 J-2 n} \Gamma \left(\frac{1}{2} (2 J+1)\right) \,
   _2F_1\left(\frac{1}{2}-\frac{J}{2},-n;\frac{1}{2}-J;-\frac{1}{3}\right) \Gamma
   \left(\frac{1}{2} (2 n-J)\right)}{\sqrt{\pi } \Gamma \left(-\frac{J}{2}\right) \Gamma  (J+1) \Gamma (n+1)}\, .
   \end{split}
    \end{equation}}}
\noindent We also have the following expression for the Feynman block $\mathcal{F}_B(n,J,\rho)$ in eq(\ref{fbclosedform}):
{\small \begin{eqnarray} \label{fbclosed}
\mathcal{F}_B(n,J,\rho)= -\mathcal{U}_2(n,J,\rho)-\mathcal{H}(n,J,\rho) \,,
\end{eqnarray}
where, 
\begin{eqnarray}
\mathcal{H}(J,n,\rho)&=&2^J \sum_{k=0}^J \binom{J}{k}\binom{\frac{J+k-1}{2}}{J} (-1)^n (-1-2 \rho)^{k/2}\nonumber\\&+& 2^J \sum_{k=0}^J \sum_{m=0}^n\sum_{l=0}^{n-m}(-1)^m 2^{2m-n}\binom{J}{k}\binom{\frac{J+k-1}{2}}{J} \binom{k}{m}\binom{n-m}{l} \left(1+(-1)^l \right) (-1-2 \rho)^{l/2}\,, \nonumber\\
\mathcal{U}_2(J,n,\rho)&=&\mathcal{U}_1(J,n,\rho)
+\sum_{k=0}^{J/2-1} \frac{2(-1)^{n+1}(1+k)\left(k+2\right)_{n-1}}{\left(2\right)_{n-1}} \mathcal{T}\left(J,J/2-1-k\right)~~{\rm where,} \nonumber\\
\mathcal{U}_1(J,n,\rho)&=& \sum_{k=0}^{J/2-1} \mathcal{S}_1(k,n,\rho) \mathcal{T}\left(J,J/2-1-k\right) ~~{\rm with}\nonumber\\
\mathcal{S}_1(k,n,\rho)&=&(-1)^n\left((-1)^{k+1} \binom{-n}{k+1}+ (1-\rho )^k (\rho +1)^n+2^{-k-1}  (1-\rho )^k
   \left(-2^{k+1} (\rho +1)^n+(\rho -1)\right) \right) \,. \nonumber
\end{eqnarray}}

%%%%%%%%%%%%%%%%%%%%%%%%%%%%%%%%%%%%%%%%%%%%%%%%%%%%%%%%%%%%%%%

\section{Ramanujan's master theorem}
\label{sec:RMT}

Consider a function $f(x)$ which can be expanded as
\begin{equation}
\begin{split}
f(x) = \sum_{k=0}^{\infty} \frac{\phi(k)}{k!} (-x)^{k}
\end{split}
\end{equation}
\noindent Then according to Ramanujan's master theorem, the Mellin transform of $f(x)$ is given by
\begin{equation}
\label{rmt}
\begin{split}
\widetilde{f}(s) = \int_{0}^{\infty} dx \ x^{s-1} \ f(x) = \Gamma(s) \phi (-s) 
\end{split}
\end{equation}
In particular applying the theorem to $g(x)=\sum_{k=0}^{\infty} \psi(k) (-x)^{k}$ i.e., for $\psi(k)=\phi(k)~ k!$ and using the reflection identity we get:
\begin{equation}
\label{rmt1}
\begin{split}
\widetilde{g}(s) = \int_{0}^{\infty} dx \ x^{s-1} \ g(x) = \frac{\pi}{\sin{\pi s}} \psi (-s) 
\end{split}
\end{equation}
which is the version we use in \eqref{mteftexp}. A few comments are in order:
\begin{itemize}
\item {\bf Behaviour around the origin:} The theorem on first glance appears to suggest that the Mellin transform is completely determined by the behaviour of a function $f(x)$ around the origin. However this is clearly not true since need a closed form of the Taylor coefficient $\phi(k)$ i.e., know $\phi(k)$ {\emph for all} $k$ to apply the theorem which is equivalent to knowing the full function.
\item {\bf Uniqueness:} The above theorem says the Mellin transform is obtained by analytically continuing the Taylor coefficient $\phi(n)$'s to the Mellin parameter $-s$. One might wonder if such an analytic continuation is unique. It seems like the answer is obviously no since  there are several analytic continuations that are possible for instance consider $\phi(n)=n!$ then $\phi(s)$ could $\Gamma(s)$ but it could also be any element of the following 1-parameter family $\Gamma(s)+ a \sin{\pi n}$. However if we put a restriction on how fast $\phi(s)$ can grow at large values of $s$ then the result is unique by Carlson's theorem \cite{hardy,moll}.
\item {\bf Existence:} One can readily check that the theorem holds for $f(x)=e^{-x}$ as $\phi(k)=1$ in this case and thus we get $\tilde{f}(s)=\Gamma(s)$. However the theorem also fails for $f(x)=0$ with $\phi(k)=(-1)^k\sin{\pi k}$ since the theorem says $\tilde{f}(s)=-\pi \neq 0$. The precise conditions and when the theorem holds were worked out by G.H.Hardy \cite{hardy}.For the theorem to be valid \cite{hardy,moll} we need  $\phi(u)$ for $u = v + i w $ to be  regular on some  strip $H(\delta)= \{u | v \ge - \delta \} $ and  there exists parameters  $A < \pi ,P \in \mathbb{R},C>0$ such that $ |a(u)| < C e^{P v + A |w| }$. A counter example is $\phi(k)=\sin{\pi k}$ which has $A=\pi$ which explains why the theorem fails in this case.
 \item {\bf For physical amplitudes:} For amplitudes that obey the Froissart bound then $\mathcal{M}(\omega^2,-z\omega^2)= o(\omega^4)$ then Mellin transform converges and is regular on strip around $Re(s)=-4$. Since the theorem depends on the growth properties the Taylor coefficients at low energies namely $\tilde{\mathcal{W}}(n,z)$ defined in \eqref{wzntilde}. Assuming $|\mathcal{W}_{p,q}|\le \lambda$ then for large $n$ we have 
 \begin{eqnarray}
    | \tilde{\mathcal{W}}(n,\rho) | &<&\lambda e^{\frac{n}{2} \log{\rho}}
\end{eqnarray}
For $0<\rho<1$ we have $P=\frac{\log{\rho}}{2} <0 , A=0 <\pi $ and for $-1<\rho<0$ we have $P=\frac{\log{|\rho|}}{2}, A=-\frac{\pi}{2}<\pi$. Thus the conditions of the theorem are satisfied.
 \item {\bf Universal factor:} The factor $\frac{\pi}{\sin{\pi s}}$ is universal also follows from the theorem directly. In \cite{Arkani-Hamed:2020gyp,Chang:2021wvv} the poles at $\beta=\pm 2n$ with $n\ge 0$ were argued to capture the IR/UV physics and since $\beta=2 s $ this is consistent with their claim.
\end{itemize}

The several variable generalization of Ramanujan's master theorem has also been useful in computing mutli-loop Feynman integrals \cite{Ananthanarayan:2021not,moll1,moll2,moll3}.

%%%%%%%%%%%%%%%%%%%%%%%%%%%%%%%%%%%%%%%%%%%%%%%%%%%%%%%%%%%%%%%

\section{Positivity of $d^{(n)}_{m}$: Further examples}
\label{sec:dposexamples}

%{\bf AS: I would still shorten this much further giving explanations of the steps in words rather than spelling out in so much details.}

Here we consider the relation \eqref{dalphatrel} and analyse the conditions under which $d^{(n)}_{m} \ge 0$ for $n=4,6,7$. The case $n=5$ has already been considered in section \ref{subsec:dnmpos}. 
\subsection*{$n=4$ :}
For $n=4$, we have
\begin{equation}
\label{dn4m}
\begin{split}
&  d^{(4)}_{m} = \sum_{J=0}^{\infty}(2J+1)  \   \chi^{(4)}_{m}(J) \ \widetilde{\alpha}_{J}(4, \delta_{0}), \quad \quad m=0,1,2
\end{split}
\end{equation}

where 
\begin{equation}
\label{chin4m}
\begin{split}
&  \chi^{(4)}_{0}(J) = \frac{1}{2} - \frac{1}{8} J(J+1)(J^{2}+J-8), \quad  \chi^{(4)}_{1}(J) = \frac{3}{2} \chi^{(4)}_{2}(J)=  - \frac{3}{8} J(J+1)(J^{2}+J-8)
\end{split}
\end{equation}

Now for $n=4$ we also have the null constraint
\begin{equation}
\label{lcn4}
\begin{split}
 \sum_{J=2}^{\infty} (2J+1) \ J(J+1)(J^{2}+J-8) \ \widetilde{\alpha}_{J}(4,\delta_{0})  = 0
 \end{split}
\end{equation}

From \eqref{chin4m}, we see that both $\chi^{(4)}_{1}(J), \chi^{(4)}_{2}(J)$ are proportional to the coefficient appearing in \eqref{lcn4}. This implies $d^{(4)}_{1}= d^{(4)}_{2}=0$. Now from the definition of $d^{(4)}_{m}$ in terms of Wilson coefficients, given by \eqref{betaresrhovar}, we already know that $d^{(4)}_{m}= 0$, for $m=1,2$. So this is a consistency check. Then let us take $m=0$ in \eqref{dn4m}. Using \eqref{chin4m} and \eqref{lcn4} we get 
\begin{equation}
\label{dn4m0a}
\begin{split}
  d^{(4)}_{0} & = \frac{1}{2}\bigg[ \widetilde{\alpha}_{0}(4, \delta_{0})   +  \sum_{J=2}^{\infty}(2J+1) \  \widetilde{\alpha}_{J}(4, \delta_{0})  \bigg] 
\end{split}
\end{equation}

Since $\widetilde{\alpha}_{J}(4,\delta_{0}) \ge 0$ due to unitarity, \eqref{dn4m0a} implies
\begin{equation}
\label{dn4m0b}
\begin{split}
  d^{(4)}_{0} \ge 0
\end{split}
\end{equation}

The fact that $d^{(4)}_{0} \ge 0$ has to hold is expected due to unitarity. In terms of the Wilson coefficients, $\mathcal{W}_{20}x^2=\omega^8 d^{(4)}_0(1-\rho)^2\ge 0$. It is known that unitarity implies $\mathcal{W}_{20}\geq 0$. However, it is interesting to note that in our analysis the condition $d^{(4)}_{0} \ge 0$ becomes manifest only after using the locality constraints as shown above.

%%%%%%%%%%%%%%%%%%%%%%%%%%%%%%%%%%%%%%%%%%%%%%%%%%%%%%%%%%%%%%%

\subsection*{$n=6$ :}

In this case we have
\begin{equation}
\label{dn6m}
\begin{split}
&  d^{(6)}_{m} =  \sum_{J=0}^{\infty}(2J+1)\  \widetilde{\alpha}_{J}(6, \delta_{0}) \   \chi^{(6)}_{m}(J), \quad \quad m=0,1,2,3
\end{split}
\end{equation}

where
\begin{equation}
\label{chin6m}
\begin{split}
&  \chi^{(6)}_{0}(J) = \frac{1}{4} + \frac{c_{1}(6,J) }{32}, \quad  \chi^{(6)}_{1}(J)  = \frac{5}{48} \ c_{1}(6,J), \\
& \chi^{(6)}_{2}(J) = \frac{11}{96} \ c_{1}(6,J) + \frac{1}{16} \left(6+ J(J-3)(J+1)(4+J) \right), \\
& \chi^{(6)}_{3}(J) = \frac{1}{24} \ c_{1}(6,J) + \frac{1}{16} \left(6+ J(J-3)(J+1)(4+J) \right)
\end{split}
\end{equation}

and 
\begin{equation}
\label{c1n6J}
\begin{split}
&   c_{1}(6,J) = -\frac{1}{24} (J-3) J (J+1) (J+4) \left(J (J+1) \left(J^2+J-32\right)+204\right)\,.
\end{split}
\end{equation}

Now $c_{1}(6,J) $ given above is precisely the coefficients that appears in the locality constraint equation for $n=6$. This leads to $d^{(6)}_{1}=0$. We can also easily check that $d^{(6)}_{2} = d^{(6)}_{3}$. This is a simple consequence of using \eqref{chin6m} and noting that in \eqref{dn6m}, the terms proportional to $c_{1}(6,J)$ drop out due to the null constraint. Thus we only need to analyse the cases $m=0$ and $m=2$.  For $m=0$, applying the locality constraint we can express $d^{(6)}_{0}$ as
\begin{equation}
\label{dn6m0}
\begin{split}
&  d^{(6)}_{0} =  \frac{1}{4} \bigg[ \widetilde{\alpha}_{0}(6, \delta_{0}) \   +   \sum_{J=2}^{\infty}(2J+1)\  \widetilde{\alpha}_{J}(6, \delta_{0})     \bigg] 
\end{split}
\end{equation}

Clearly $d^{(6)}_{0} \ge 0$ since $\widetilde{\alpha}_{J}(6,\delta_{0}) \ge 0$. Similarly for $m=2$ we obtain
\begin{equation}
\label{dn6m2}
\begin{split}
&  d^{(6)}_{2} =  \frac{1}{8} \bigg[ 3\widetilde{\alpha}_{0}(6, \delta_{0})   - 75 \widetilde{\alpha}_{2}(6, \delta_{0}) +  \frac{1}{2} \sum_{J=4}^{\infty}(2J+1) \left( 6+J(J-3)(J+1)(4+J) \right) \  \widetilde{\alpha}_{J}(6, \delta_{0})     \bigg] 
\end{split}
\end{equation}

Therefore, for $d^{(6)}_{2}\ge 0$ to hold, the sufficient condition is 
\begin{equation}
\label{dn6m2b}
\begin{split}
\widetilde{\alpha}_{0}(6, \delta_{0})  \ge   25 \hspace{0.05cm} \widetilde{\alpha}_{2}(6, \delta_{0})
\end{split}
\end{equation}
\subsection*{$n=7$ :} 
Finally let us consider the case $n=7$,
\begin{equation}
\label{dn7m}
\begin{split}
&  d^{(7)}_{m} = \sum_{J=0}^{\infty}(2J+1)\  \widetilde{\alpha}_{J}(7, \delta_{0}) \   \chi^{(7)}_{m}(J), \quad \quad m=0,1,2,3
\end{split}
\end{equation}

where
\begin{equation}
\label{chin7m}
\begin{split}
&  \chi^{(7)}_{0}(J) = -\frac{1}{160}\left( 10 c_{1}(7,J) - 3c_{2}(7,J) \right) \\
& \chi^{(7)}_{1}(J) = -\frac{13}{320} c_{2}(7,J) + \frac{7}{8} +\frac{1}{72}J(J+1)[276+J(J+1)(2J(J+1)-61)] \\
& \chi^{(7)}_{2}(J) = -\frac{3}{8} c_{1}(7,J) + \frac{23}{160} c_{2}(7,J), \quad  \chi^{(7)}_{3}(J) = -\frac{3}{16} c_{1}(7,J) + \frac{5}{64} c_{2}(7,J) 
\end{split}
\end{equation}
and 
\begin{equation}
\label{c1n7J}
\begin{split}
&   c_{1}(7,J) = \frac{1}{720} J (J+1) [J (J+1) (J (J+1) (J (J+1) (2 J (J+1)-155)+4836)-65468)+235200] \\
& c_{2}(7,J) = \frac{1}{360} J (J+1) [J (J+1) (J (J+1) (J (J+1) (2 J (J+1)-155)+4916)-67908)+246960]
\end{split}
\end{equation}
The coefficients  $c_{1}(7,J), c_{2}(7,J)$ are identical to the ones that enter in the two independent locality constraint equations for $n=7$. Thus we straightforwardly get $d^{(7)}_{0} = d^{(7)}_{2} = d^{(7)}_{3} =0$. 

Now for $m=1$, applying the locality constraints, it can be easily shown that $d^{(7)}_{1}$ becomes
\begin{equation}
\label{dn7m1c}
\begin{split}
  d^{(7)}_{1} & =  4 \bigg[ 7 \hspace{0.05cm} \widetilde{\alpha}_{0}(7, \delta_{0}) -  75 \hspace{0.05cm} \widetilde{\alpha}_{2}(7, \delta_{0})  + \frac{1}{60480} \sum_{J=8}^{\infty}(2J+1) (J-6) (J-4) (J+5) (J+7) \big[J (J+1) \\
  & (7 J (J+1) (2 J (J+1)-31)+414)+504\big] \ \widetilde{\alpha}_{J}(7, \delta_{0})  \bigg] 
\end{split}
\end{equation}
This implies that for $d^{(7)}_{1}$ to be non-negative it is sufficient to have
\begin{equation}
\label{dn7m1d}
\begin{split}
     \widetilde{\alpha}_{0}(7, \delta_{0}) \ge   \frac{75}{7} \hspace{0.05cm} \widetilde{\alpha}_{2}(7, \delta_{0})
\end{split}
\end{equation}
In a similar manner one also can show that a sufficient condition for $d_m^{(8)}\geq 0$ is
\begin{equation}
\label{dn8m1d}
\begin{split}
     \widetilde{\alpha}_{0}(8, \delta_{0}) \ge   13.75 \hspace{0.05cm} \widetilde{\alpha}_{2}(8, \delta_{0})
\end{split}
\end{equation}
For convenience we repeat the sufficient conditions for $d_m^{(i)}\ge 0$, for $i=5,6,7,8$ below:
\begin{eqnarray}
    \widetilde{\alpha}_{0}(5, \delta_{0}) &\ge&   11.2 \hspace{0.05cm} \widetilde{\alpha}_{2}(5, \delta_{0})\,, \quad \widetilde{\alpha}_{0}(6, \delta_{0}) ~~\ge~~   25 \hspace{0.05cm} \widetilde{\alpha}_{2}(6, \delta_{0})\,,\\ \widetilde{\alpha}_{0}(7, \delta_{0}) &\ge&   10.72 \hspace{0.05cm} \widetilde{\alpha}_{2}(7, \delta_{0})\,,\quad \widetilde{\alpha}_{0}(8, \delta_{0}) ~\ge~   13.75 \hspace{0.05cm} \widetilde{\alpha}_{2}(8, \delta_{0})\,.
\end{eqnarray}

Let us also note a simple consequence of the properties of partial wave moments, without assuming $d_m^{(n)}$ positivity. Using \eqref{dn4m0a} and \eqref{dn5m1e}, we find
\begin{equation}
   \left (d_0^{(4)}-\frac{2}{5}d_1^{(5)}\right)=\widetilde\alpha_0(4,\delta_0)-\widetilde\alpha_0(5,\delta_0)+ ({\rm positive~terms})\ge 0\implies d_0^{(4)}\ge \frac{2}{5} d_1^{(5)}\,.
\end{equation}

Here we have used the fact that for moments $\widetilde \alpha_0(4,\delta_0)\ge \widetilde\alpha_0(5,\delta_0)$.
In terms of the $\mathcal{W}_{p,q}$'s this translates to 
\begin{equation}\label{w20w11}
    \mathcal{W}_{20}\ge -\frac{2}{5}\mathcal{W}_{11}\,.
\end{equation}
In type-II string theory, $lhs\approx 2.017 > rhs\approx 1.994$.

%%%%%%%%%%%%%%%%%%%%%%%%%%%%%%%%%%%%%%%%%%%%%%%%%%%%%%%%%%%%%%

\section{String theory example: Details}
\label{sec:stringexamples}

Here we spell out the type-II string theory details which can be used to cross-check many of the inequalities proved in the main draft. Note that although the amplitude is in 10 spacetime dimensions, the $\tilde \alpha_J$-moments are positive using Legendre polynomials. This is all that we will need for all our checks. Let us focus on the Gamma function dependence on the amplitude (the full amplitude is multiplied by $x^2$ which one can factor out for convenience as well as better convergence). The amplitude is given by 
\begin{equation}
   {\mathcal M}(s,t)=- \frac{\Gamma(-s)\Gamma(-t)\Gamma(-u)}{\Gamma(1+s)\Gamma(1+t)\Gamma(1+u)}\,,\quad s+t+u=0\,.
\end{equation}
Using the Celestial variables and expanding around $\omega^2=0$ gives eq.(\ref{cbtype2}). The partial wave coefficients $\alpha_J(s)$ are given by
\begin{equation}
\alpha_J(s)=\frac{\delta(s-n)}{32}\int_{-1}^1 dx\, (-1)^{n+1}\frac{\Gamma(n+\frac{n}{2}(x-1))\Gamma(-\frac{n}{2}(x-1))}{2(n!)^2 \Gamma(1-n-\frac{n}{2}(x-1))\Gamma(1+\frac{n}{2}(x-1))}P_J(x)\,.    
\end{equation}
Plugging this into eq.(\ref{fbex}) and truncating to $J_{max}=20, n_{max}=10$ leads to $\approx 0.3\%$ agreement with the rhs of eq.(\ref{cbtype2}). The agreement improves with increasing $J_{max}, n_{max}$ as expected.

%%%%%%%%%%%%%%%%%%%%%%%%%%%%%%%%%%%%%%%%%%%%%%%%%%%%%%%%%%%%%%%
For convenience and ready reference, we tabulate some of the partial wave moments, $\tilde\alpha_J(n,1)$ as defined in eq.(\ref{pwmoments}), obtained using $n_{max}=50$:
\begin{center}
\begin{tabular}{|c|c|c|c|c|c|}
\hline
$n$\textbackslash $J$ & 0 & 2 & 4 & 6 & 8\\
\hline
1 & 1.0237 & 0.01066 & 0.00067 & 0.000095 & 0.000021\\
2 & 1.0113 & 0.00472 & 0.00019 & 0.000019 & 3.021$\times 10^{-6}$\\
3 & 1.0055 & 0.00226 & 0.00006 & 3.95 $\times 10^{-6}$ & 4.78 $\times 10^{-7}$\\
4 & 1.0027 & 0.00110 & 0.00002 & 8.72 $\times 10^{-7}$ & 8.00$\times 10^{-8}$\\
5 & 1.0013 & 0.00054 & 5.62 $\times 10^{-6}$ & 1.99 $\times 10^{-7}$ & 1.39 $\times 10^{-8}$\\
6 & 1.0007 & 0.00026 & 1.81 $\times 10^{-6}$ & 4.63 $\times 10^{-8}$ & 2.50 $\times 10^{-9}$\\
\hline

\end{tabular}

\end{center}
We also list the low energy Wilson coeffcients $\mathcal{W}_{p,q}$ of the II-string amplitude in the table below:
\begin{center}
\begin{tabular}{|c|c|c|c|c|}
\hline
$\mathcal{W}_{p,q}$ & $q=0$ & $q=1$ & $q=2$ & $q=3$ \\
\hline
$p=0$ & 2.40411 & -2.88988 & 2.98387 & -2.99786 \\
\hline
$p=1$& 2.07386& -4.98578 & 7.99419 & -10.9987 \\
\hline
$p=2$ & 2.0167 & -6.99881 & 14.9984 & -25.9995 \\
\hline
$p=3$ & 2.00402 & -9.00023 & 23.9996 & -49.9998\\
\hline

\end{tabular}

\end{center}
\vskip 4pt

\typeout{}

\end{document}